\documentclass{article}

\usepackage{geometry}
\usepackage{authblk}
\usepackage{amsmath}
\usepackage{graphicx}
\usepackage{amssymb}
\usepackage{algorithmicx}
\usepackage{algpseudocode}
\usepackage{listings}
\usepackage{caption}
\usepackage{subcaption}
\usepackage{hhline}

\usepackage[table,usenames,dvipsnames]{xcolor}
\usepackage{tabularx}
\usepackage{multirow}
\usepackage{url}
\begin{document}
\title{Fast Distance Fields for Fluid Dynamics Mesh Generation on Graphics Hardware}
\date{March, 2019}
\author{A. Roosing \thanks{Corresponding author ar694@cam.ac.uk}} 
\author{O. T. Strickson} 
\author{N. Nikiforakis}
\affil{Laboratory for Scientific Computing, Cavendish Laboratory, Department of Physics, University of Cambridge}

\maketitle

\begin{abstract}
We present a CUDA accelerated implementation of the Characteristic/Scan Conversion algorithm to generate narrow band signed distance fields in logically Cartesian grids. We outline an approach of task and data management on GPUs based on an input of a closed triangulated surface with the aim of reducing pre-processing and mesh-generation times. The work demonstrates a fast signed distance field generation of triangulated surfaces with tens of thousands to several million features in high resolution domains. We present improvements to the robustness of the original algorithm and an overview of handling geometric data.
\end{abstract}

\section{Introduction}
\label{sec1}
 
Signed distance fields (SDF) find uses in domains from computer graphics\ \cite{bridson} to numerical modelling\ \cite{fedkiw}. Determining the location of explicit or implicit surfaces in grids or generating meshes to describe objects is an area of active research in many computational paradigms. Triangulated surfaces are a popular working medium and the Stereolithography (STL) file format finds wide use in areas such as CFD\ \cite{janssen} and $3$D printing\ \cite{wong}. The quick generation of  robust signed distance fields from triangulated surfaces is then of great interest to many industries and academic disciplines.

\begin{figure*}[h]
\centering
\begin{subfigure}{0.45\textwidth}
\includegraphics[width=\textwidth]{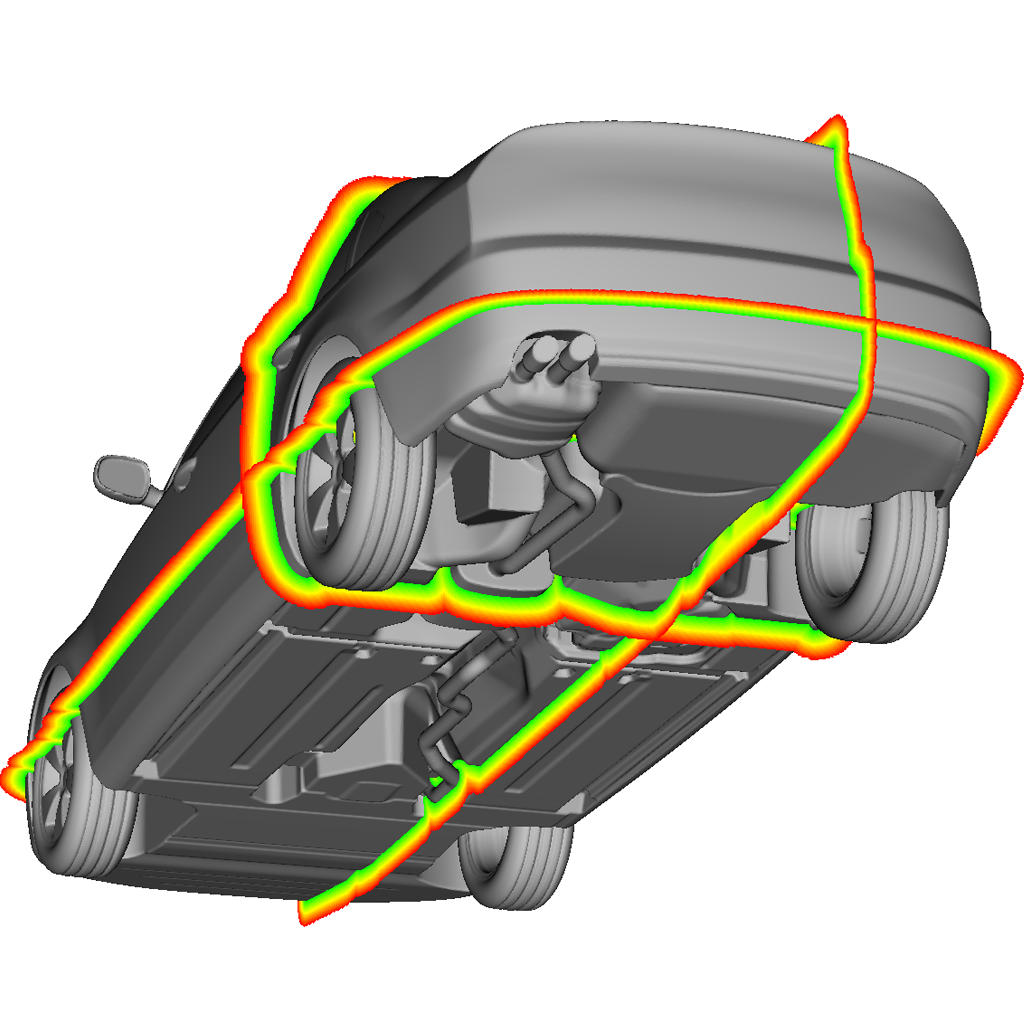}
\end{subfigure}%
\,
\begin{subfigure}{0.45\textwidth}
\includegraphics[width=\textwidth]{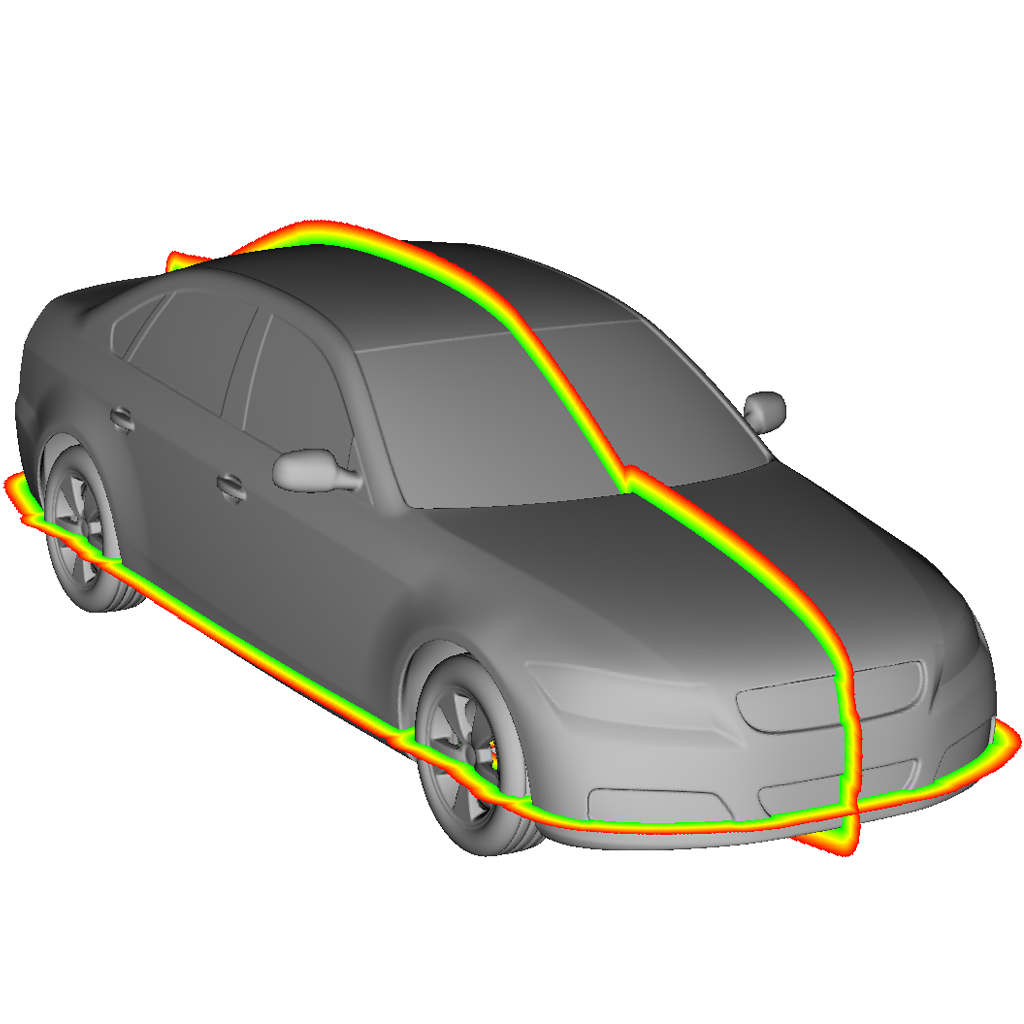}
\end{subfigure}%
\caption{The produced narrow band signed distance field and resulting surface plot of the DrivAer car model\ \cite{drivaer}. Complex geometries can be processed quickly to generate embedded meshes in the initial phases of fluid simulations. Only creating a small shell around the underlying STL model is sufficient to describe a surface intersecting a Cartesian mesh. The large number of surface features are ordered and used to build extrusions which limit the space where distance calculations are made. Due to the short run times, high resolution computational domains can be used in conjunction with detailed models, resulting in sophisticated CFD meshes with regular memory layout.}
\label{fig:car}
\end{figure*}

Often it is necessary to know only the distance to the surface within a small region around the geometry and narrow band SDFs are useful for quickly generating just the intersection between a computational mesh and an object. This finds application in embedded boundary methods in computational fluid dynamics where generating object data often takes a significant portion of the simulation set up time, which can become a bottleneck in fast prototyping when the subsequent numerical work is highly optimised and run on many-core architectures. For example, the signed distance field of a complex car body, as shown in figure \ref{fig:car}, can be used to generate cut cells in a regular computational mesh to impose boundary conditions along a detailed perimeter without introducing significant mesh generation overhead or complex connectivity information.  We focus on the generation of narrow band signed distance fields inside Cartesian grids but the algorithm discussed in this paper is potentially extendible to other paradigms.

Our main aim is to describe a robust algorithm to speedup the generation of level sets from triangulated surface information using graphics processing units (GPUs). In this paper we discuss the implementation and adjustment of the Characteristic/Scan Conversion (CSC) algorithm originally described by Mauch\ \cite{mauch}. We will outline improvements to the original approach and present an implementation on GPUs with a focus on how to manage information about many thousands of connected features. 

Park et al.\ \cite{park} have developed an algorithm for generating signed distances on the GPU for hierarchical grids. They sample mesh cells based on the complexity of the surface geometry and present a good speedup compared to identical approaches on the CPU. Their use of angle-weighted pseudonormals at surface discontinuities is similar to the strategy we employ.

Sud et al.\ \cite{sud} describe a GPU signed distance field method based on Voronoi cells and slicing. Their speedup stems from the use of GPUs, culling far away features and clamping the rasterisation of the Voronoi cells. Though their approach is different from ours, the strategy of reducing calculations is similar to the current work. Their method does not store information about the connectivity of triangles and uses the CSC algorithm for suitable sub-problems, developing a new approach for problematic surface configurations. Our implementation is purely CSC based and addresses many of these geometric cases.

Sigg et al.\ \cite{sigg} present a GPU implementation of the CSC algorithm for triangulated surfaces. Their work is focused on overcoming the need for vertex extrusions by combining edge and face extrusions. This is done in order to reduce the workload as well as avoid topological cases which the CSC algorithm finds problematic. Below we discuss a different methodology for the issues arising at vertices.

An implementation of the CSC algorithm also exists by Mauch\ \cite{bitbucket}. We use some of the insights of that code but have developed an independent strategy with updated feature generation, a high degree of parallelism and algorithmic improvements.

There is a lack of discussion in existing literature about how to best organise STL features for use with the CSC algorithm on GPUs. Specifically, it is not immediately clear how to efficiently produce extrusions from nearby surface triangles when no strict feature order is imposed in the input file. There are also gaps in the literature when it comes to discussing some complex cases that can arise in common geometries such as saddle vertices and other configurations discussed below. The main contributions of this paper are describing the efficient handling of STL features on GPUs, showing robust extrusion building for previously unaddressed surface configurations and demonstrating fast narrow band SDF generation for a variety of complex test geometries.


\section{Closest point distance transform}
\label{sec2}
\begin{figure*}[h]
\centering
\begin{subfigure}[h]{0.45\textwidth}
\includegraphics[width=\textwidth]{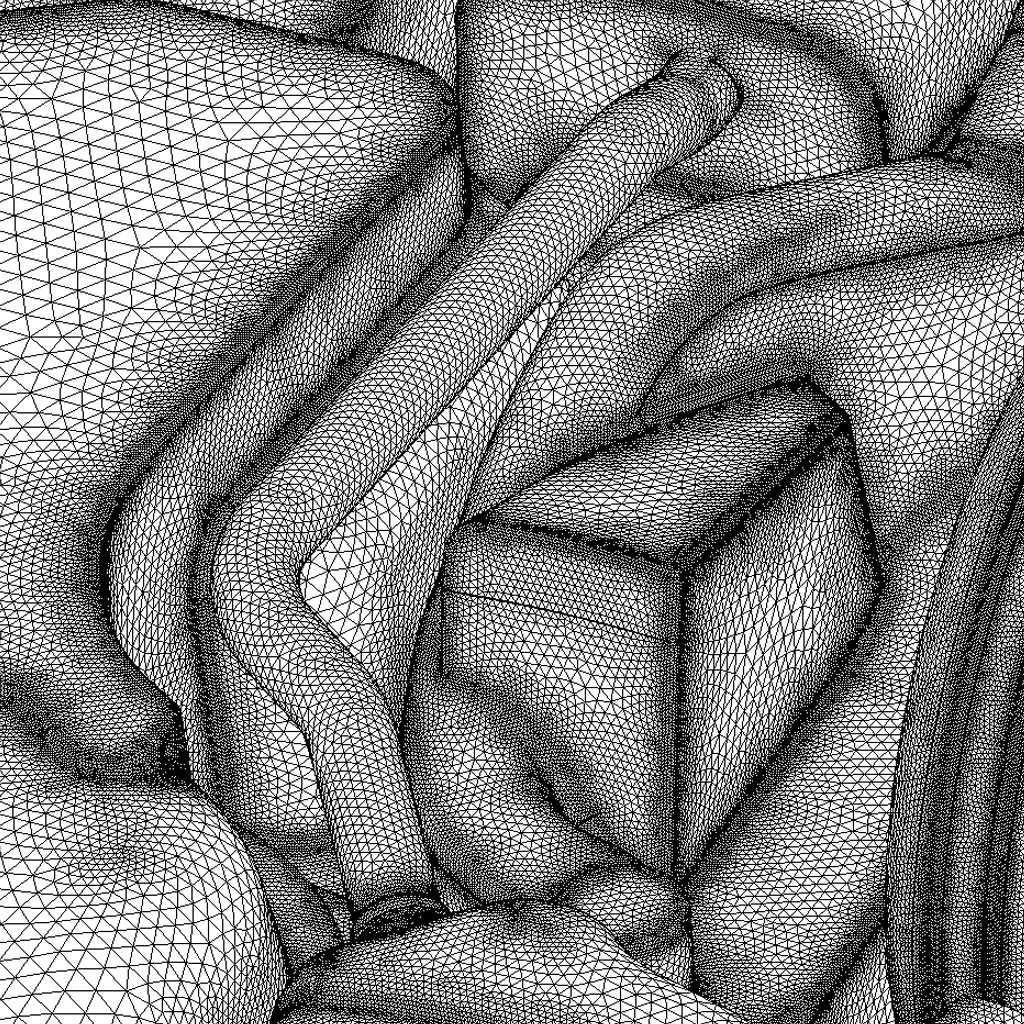}
\subcaption{Detail of DrivAer STL file}
\label{fig:detail_stl}
\end{subfigure}%
\,
\begin{subfigure}[h]{0.45\textwidth}
\includegraphics[width=\textwidth]{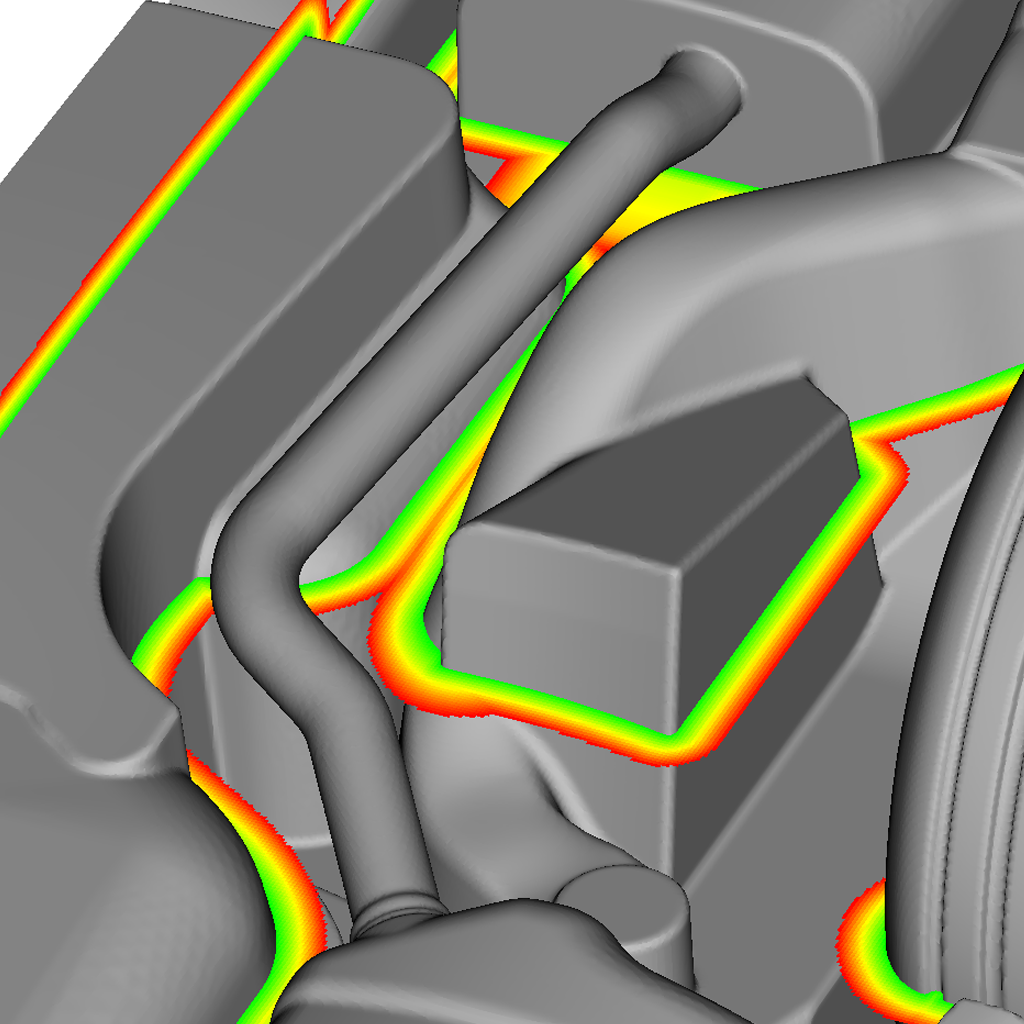}
\subcaption{Produced surface plot and SDF slices}
\label{fig:detail_sdf}
\end{subfigure}%
\caption{Narrow band SDF results for the DrivAer geometry. The quickly generated field extends to a limited distance from the surface. As the SDF is free from gaps, the $0$ crossing matches the STL input to within a fraction of $\Delta x$.}
\label{fig:detail}
\end{figure*}

The closest point distance transform algorithm\ \cite{mauch} aims to populate domain cells in the immediate vicinity of a geometry with the shortest distance to its surface. This is done by generating individual fields from triangulated surface features and combining them into a global signed distance field. Figure \ref{fig:detail} shows the input and output of the algorithm. The initial data is a collection of triangles in $3$D which describe a discontinuous surface (figure \ref{fig:detail_stl}). In a target Cartesian grid, the CSC algorithm populates the cells in the vicinity of the surface with the smallest distance to the object leading to an implicit description of the geometry (figure \ref{fig:detail_sdf}). 

The CSC algorithm can be used to generate the exact signed distance function of a surface within a regular grid. This function is defined at every point in the vicinity of the surface and grows in magnitude in the direction of the normals of the surface. For orientable surfaces, the positive and negative values divide a domain into the interior and exterior of the surface, with the surface itself lying at $0$. Let \textbf{x} be a point in the domain $\mathbb{R}^n$ and let $\partial\Omega$ be the surface. A signed distance function $f$ is then defined as:

\begin{equation}
f(\textbf{x}) = \text{min}\lbrace\text{d}\lbrace\textbf{x}, \partial\Omega\rbrace\rbrace,\ \forall \textbf{x} \in \mathbb{R}^n, 
\end{equation}
where $\text{d}\lbrace\rbrace$ gives the distance between a point and the surface.

For smooth surfaces, $f(\textbf{x})$ satisfies the Eikonal equation 

\begin{equation}
|\nabla f| = 1.
\end{equation}
In the case of discretised surfaces, however, there are discontinuities at the boundaries of the surface features. In this case, the signed distance field of the surface is the sum of the signed distance fields of all smooth regions of the surface.

The CSC algorithm uses the features of discrete surfaces to generate extrusions in their normal direction that are guaranteed to include at least the closest points to the original features. These extrusions are similar to Voronoi cells with the difference that they may include more than the closest points to a feature, they are artificially enlarged and may overlap. The sum of these extrusions will include all the closest points to the surface.

Let $d_{ijk}$ be the minimum distance from the mesh cell $c_{ijk}$ to the surface. By constructing extrusions $E$ for all of the features of the surface, the CSC algorithm can be written as:
\\
\\
\begin{minipage}{\textwidth}
\begin{algorithmic}
\State \{ $d_{ijk} = \infty$ for all $i, j ,k$ \}
\ForAll{$e \in E$} 
	\ForAll{$c_{ijk} \in e$} 
		\State $d_{\text{new}} =$ distance to feature
		\If {$|d_{\text{new}}| < |d_{ijk}|$}
			\State $d_{ijk} = d_{\text{new}}$
		\EndIf
	\EndFor
\EndFor
\end{algorithmic}
\end{minipage}
\\
\\

Calculating the minimum distances to the surface from the mesh cells within all of these areas produces a signed distance field. This operation is called a distance transform and results in an implicit description of the surface within the rectilinear mesh. As the work done is bounded by the number of surface features and the number of cells in the extrusions, the computational complexity of the algorithm is optimal: linear in both the feature count and the resolution of the mesh.

\begin{figure}[h]
\centering
\begin{subfigure}[b]{0.25\textwidth}
\includegraphics[width=\textwidth]{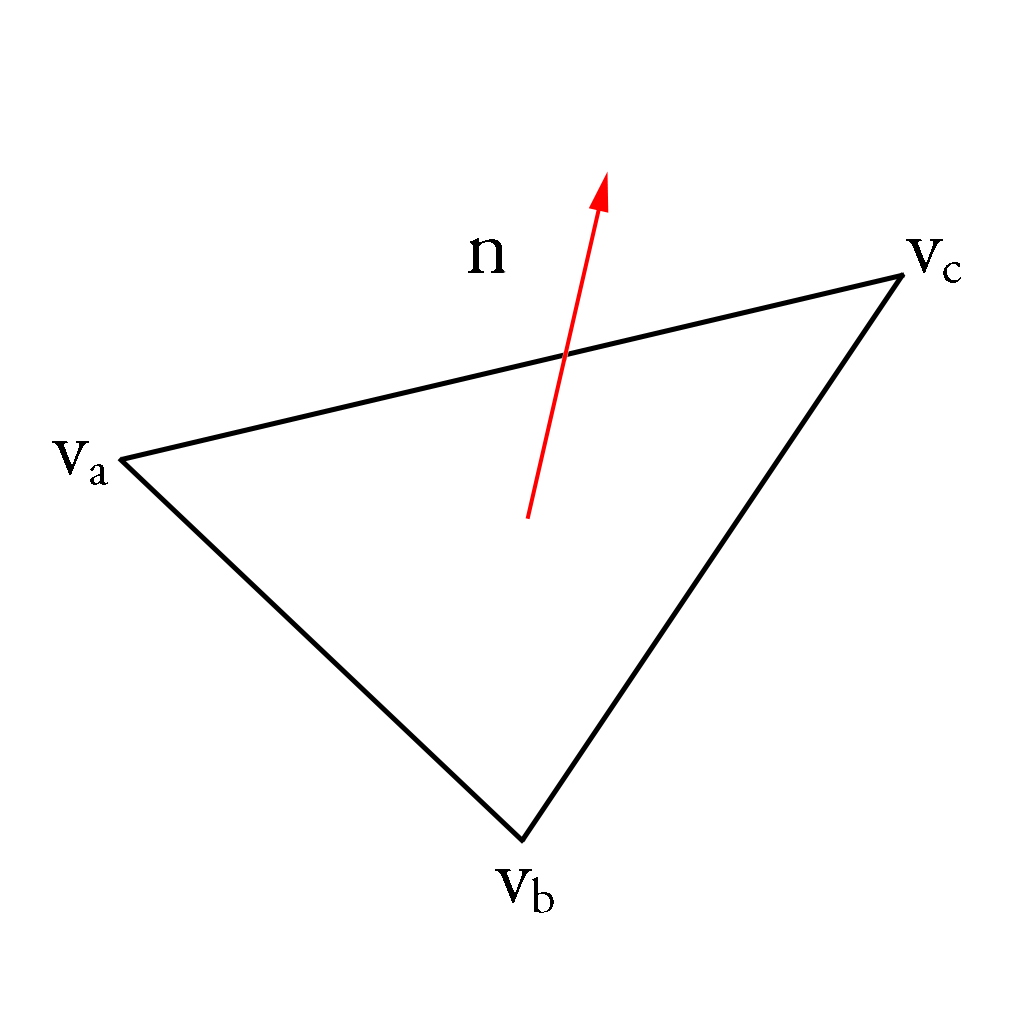}
\subcaption{STL feature}
\end{subfigure}%
~\begin{subfigure}[b]{0.25\textwidth}
\includegraphics[width=\textwidth]{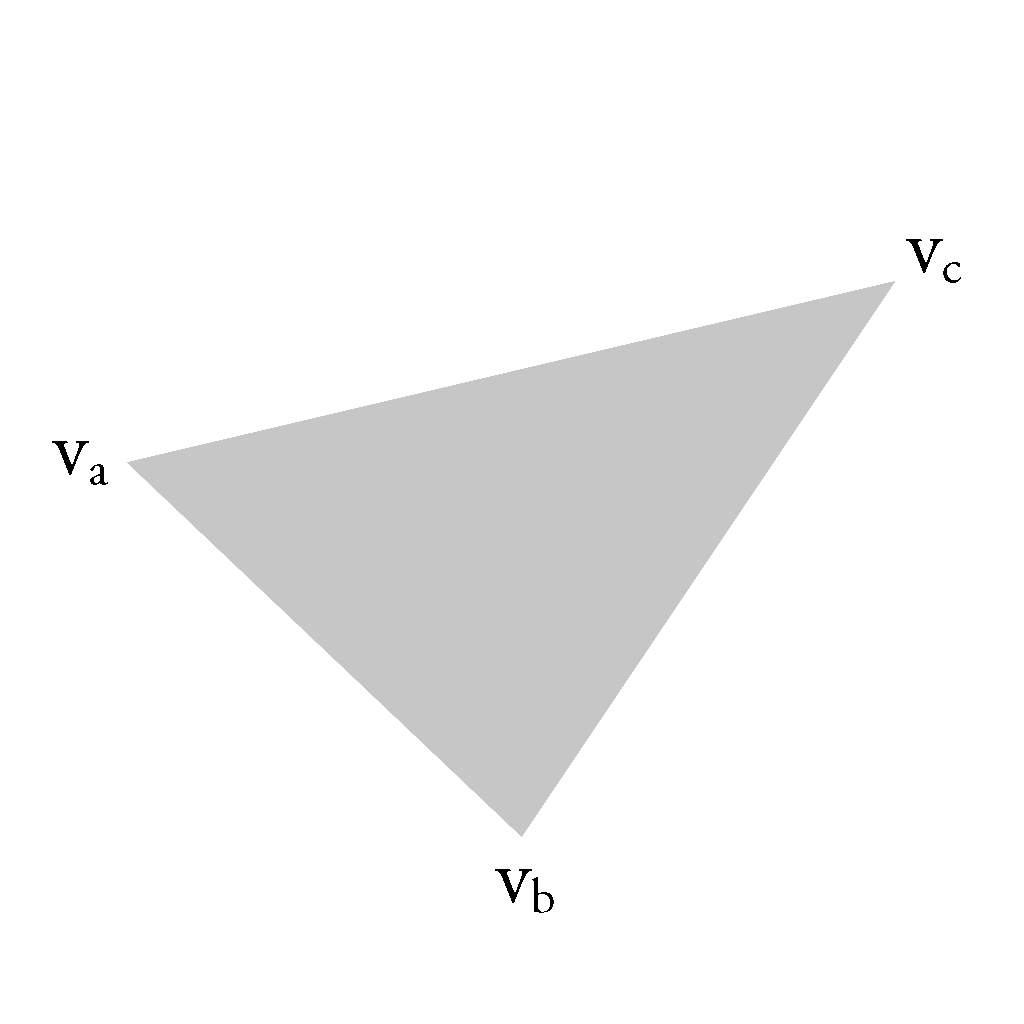}
\subcaption{face}
\end{subfigure}%
\begin{subfigure}[b]{0.25\textwidth}
\includegraphics[width=\textwidth]{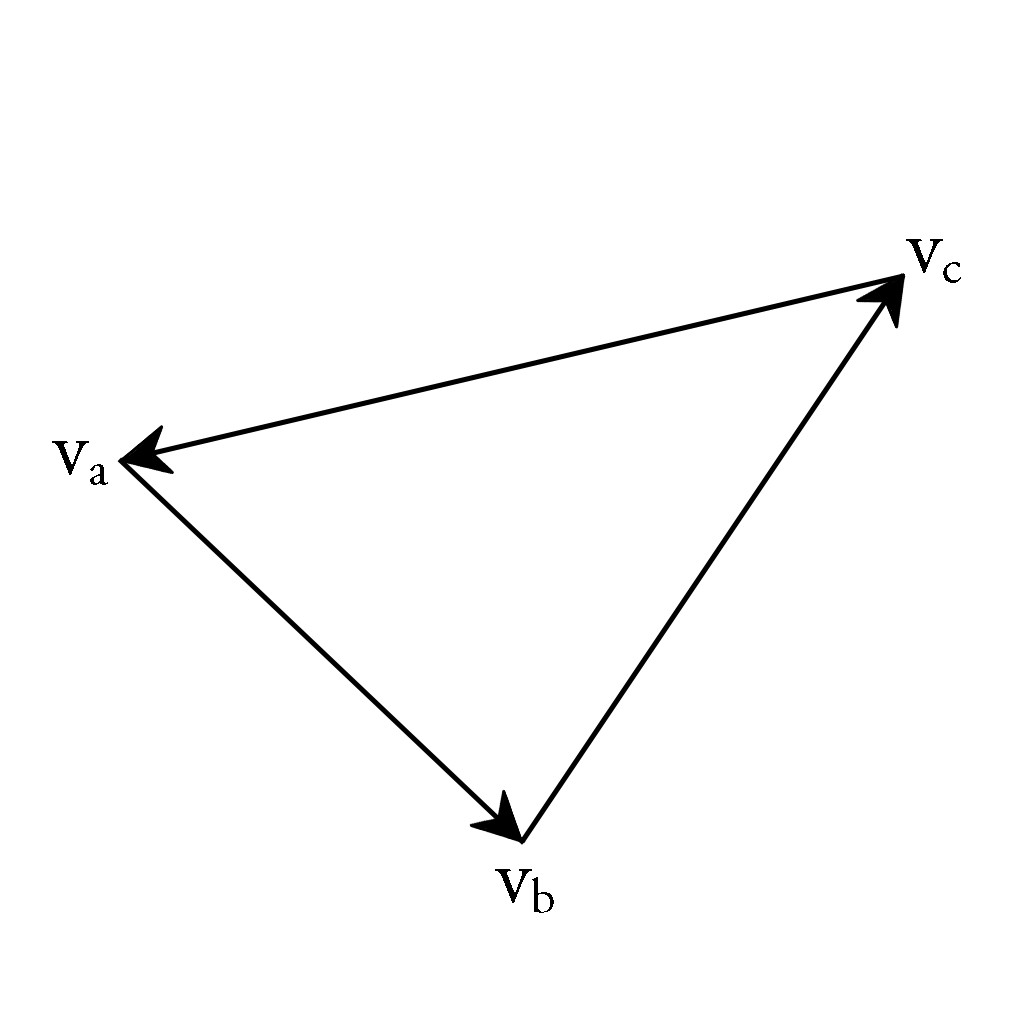}
\subcaption{edges}
\end{subfigure}%
~\begin{subfigure}[b]{0.25\textwidth}
\includegraphics[width=\textwidth]{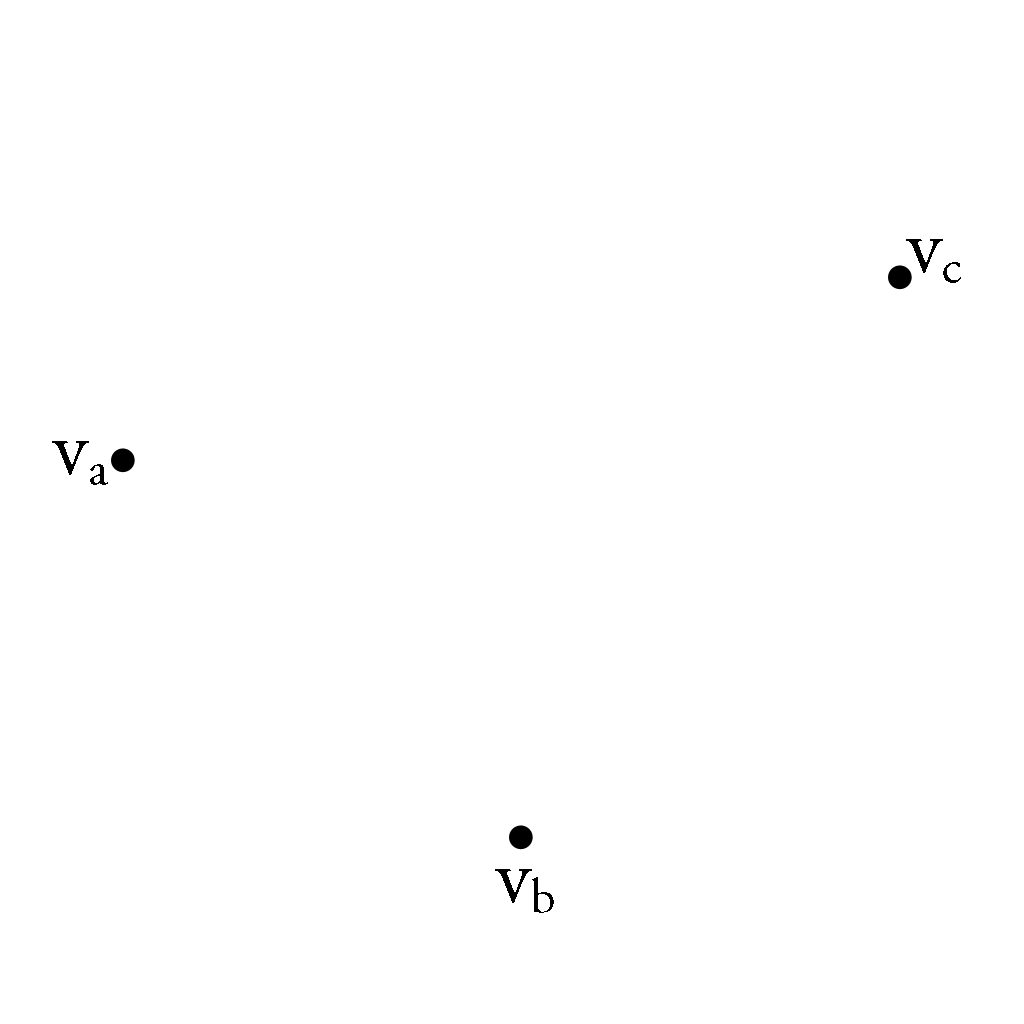}
\subcaption{vertices}
\end{subfigure}%
\caption{STL surface features. The file elements are divided into three aspects and each feature is used to generate extrusions where the closest distance to the surface lies on the feature.\label{fig:features}}
\end{figure}

The CSC algorithm is limited to orientable closed surfaces. These geometries have a well-defined interior allowing for a signed distance field where the positive and negative distances are on either side of the surface, which lies at the $0$ level set. The current work is concerned with closed triangulated surfaces in $3$D. The features of these surfaces are the triangular faces, the triangle vertices and the triangle edges as shown in figure \ref{fig:features}.

\section{Extrusions}
\label{sec3}

The extrusion polyhedra from the surface features encompass the area where the SDF is calculated. We list the different extrusion types, how they are generated and how our implementation diverges from the original description. We describe the categorisation of surface features and how this is used to reduce the amount of calculation that needs to be done, discussing unaddressed scenarios and proposed improvements.

\begin{figure}[h]
\centering
\begin{subfigure}[b]{0.25\textwidth}
\includegraphics[width=\textwidth]{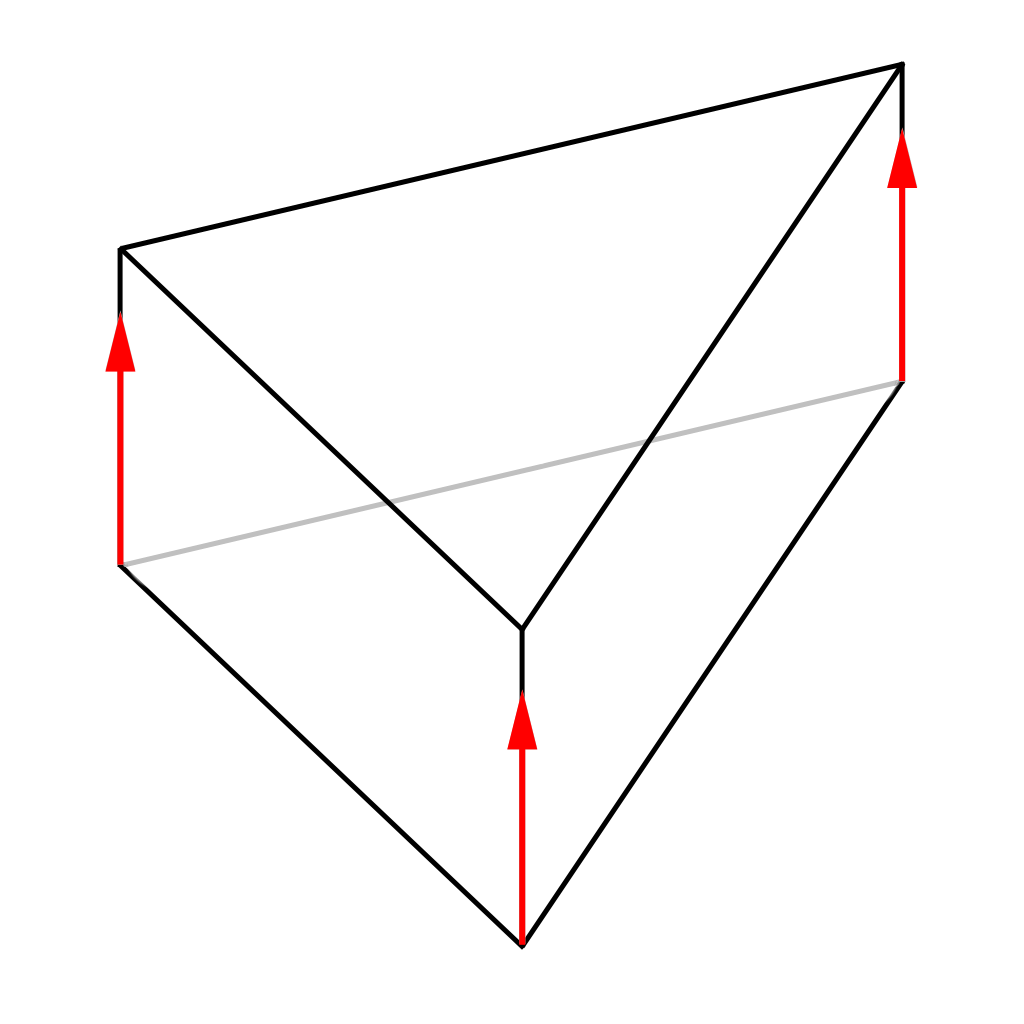}
\subcaption{face extrusion}
\label{fig:extrusions_a}
\end{subfigure}%
~\begin{subfigure}[b]{0.25\textwidth}
\includegraphics[width=\textwidth]{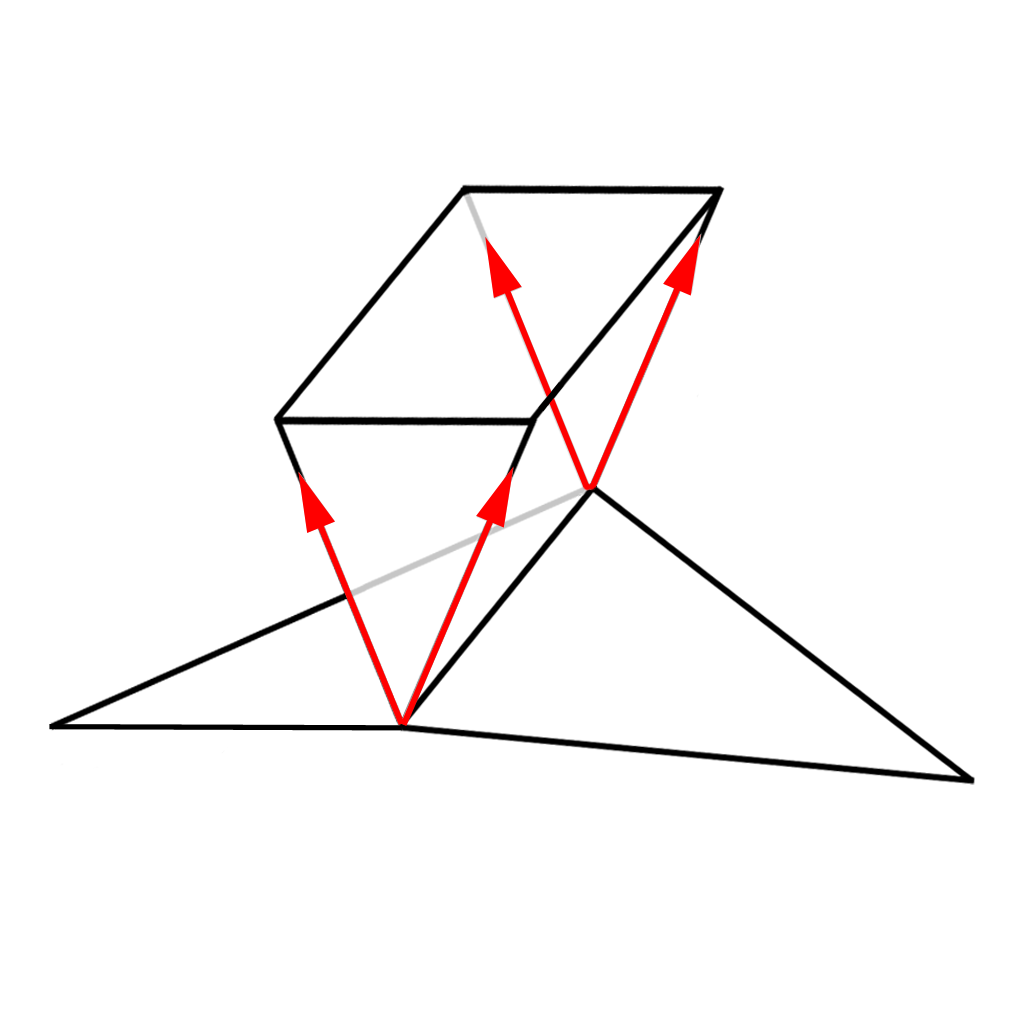}
\subcaption{edge extrusion}
\label{fig:extrusions_b}
\end{subfigure}%
\begin{subfigure}[b]{0.25\textwidth}
\includegraphics[width=\textwidth]{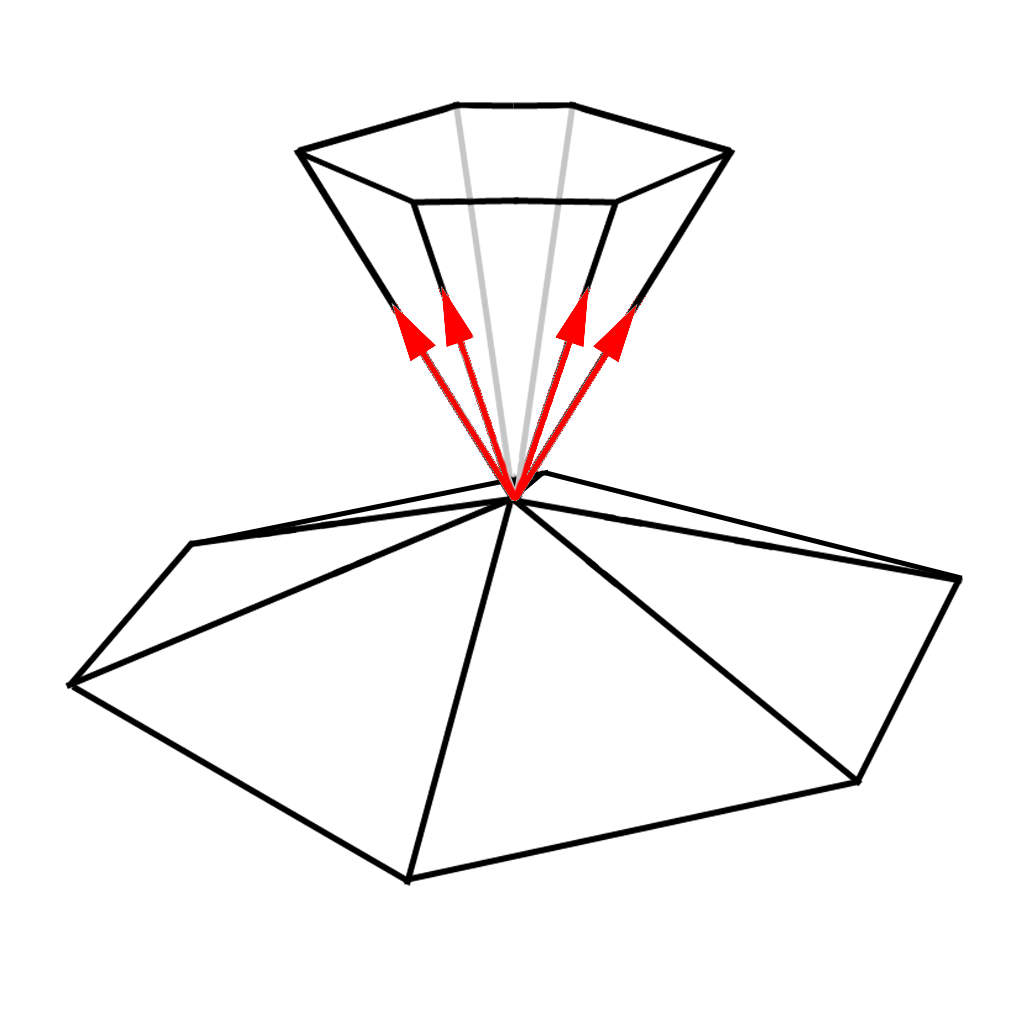}
\subcaption{vertex extrusion}
\label{fig:extrusions_c}
\end{subfigure}%
\begin{subfigure}[b]{0.25\textwidth}
\includegraphics[width=\textwidth]{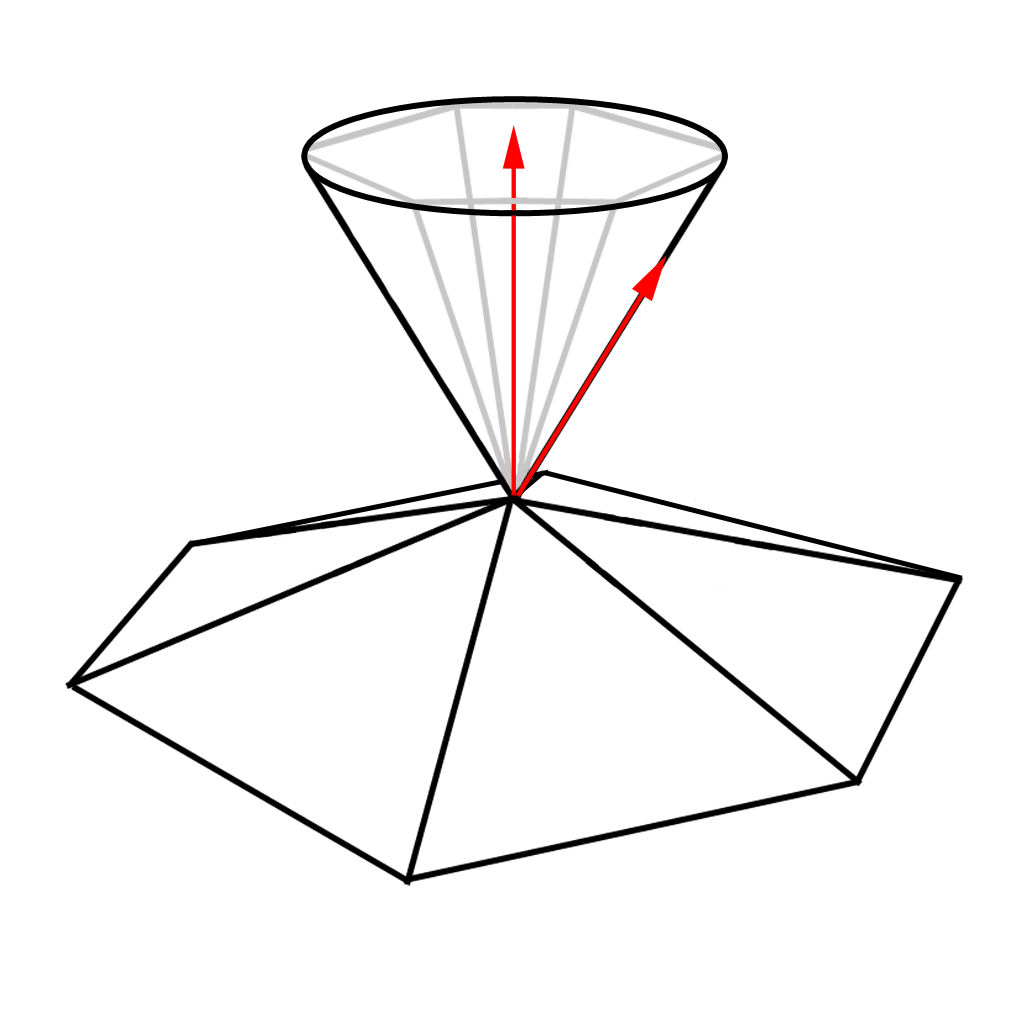}
\subcaption{vertex cone}
\label{fig:extrusions_d}
\end{subfigure}%
\caption{The extrusions from different surface features are generated in the face normal directions. The vertex extrusions can be simplified by assigning a cone which encompasses all of the face normals meeting at the vertex.}
\label{fig:extrusions}
\end{figure}

The CSC algorithm describes the construction of extrusions containing at least the closest points to the discrete features. These extrusions are constructed based on the position, limits and normals of the underlying geometries. Extruding outward from a face produces a prism in the normal direction (figure \ref{fig:extrusions_a}). An edge extrusion is a prism extruded from the line between two vertices in the directions of the two neighbouring faces (figure \ref{fig:extrusions_b}) and a vertex extrusion is a pyramid defined by the normals of the adjacent faces that meet at the vertex (figure \ref{fig:extrusions_c}). 

\begin{figure}[h]
\centering
\includegraphics[width=0.25\textwidth]{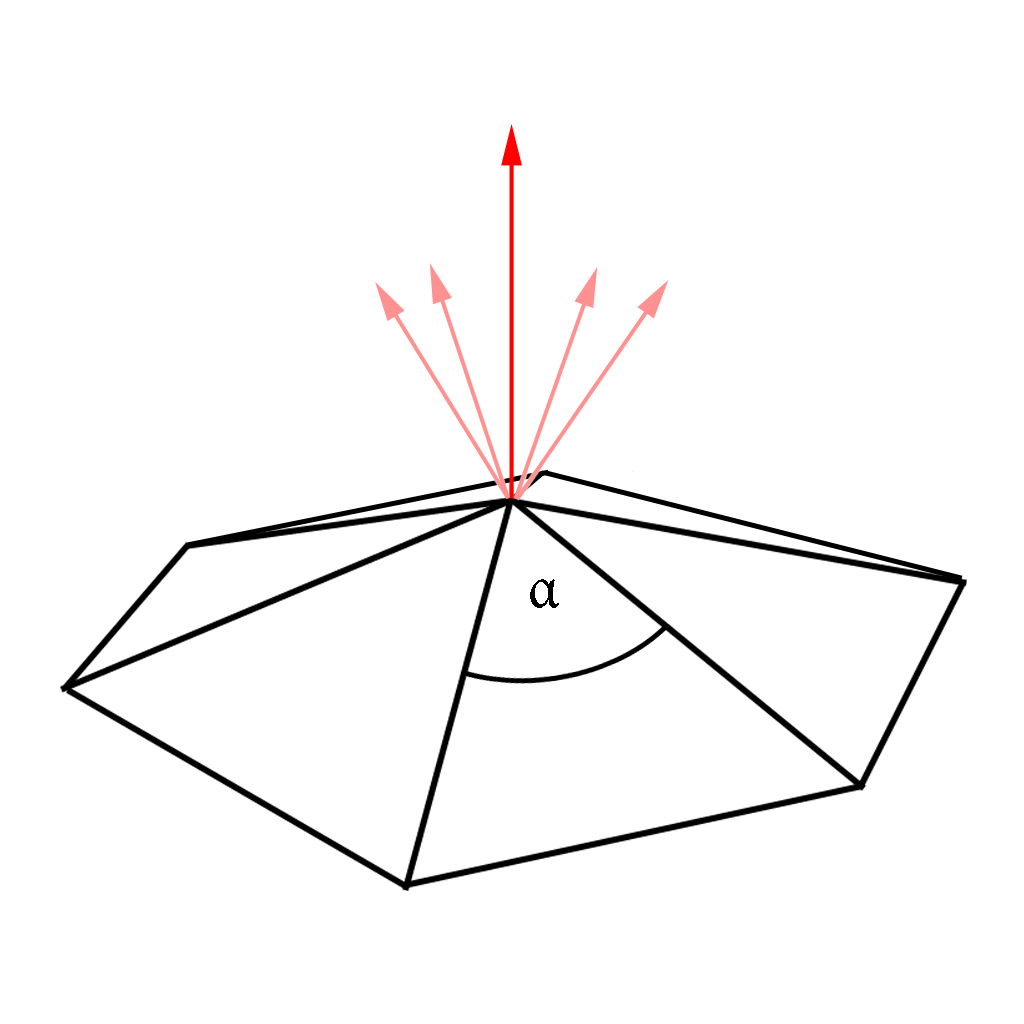}
\caption{The average pseudonormal, which is used as the axis of the cone, is constructed by weighting face normals by their respective angles $\alpha$. The weighting deals with issues arising from many coplanar faces skewing the average normal.}
\label{fig:pseudonormal}
\end{figure}

While the two prism extrusions have a known number of faces, the vertex pyramid can be of arbitrary complexity which makes implementation and workload assessment difficult. We simplify the vertex extrusion by using a cone that encompasses all of the normals of the faces that meet at the vertex (figure \ref{fig:extrusions_d}). The new vertex extrusion is constructed in the average direction of the normals weighted by the angle between the two edges of each triangle that meet at the vertex in question. Taking the unweighted average can lead to incorrect extrusions. As described by B\ae rentzen and Aan\ae s\ \cite{pseudonormal}, many coplanar faces sharing a vertex can shift the average disproportionally away from what would be the intuitive direction of the vertex. The result is an angle-weighted pseudonormal that correctly points in the average direction of the vertex (figure \ref{fig:pseudonormal}). This direction will be the axis of the cone and the adjacent normal most diverging from the average lies on the side of the cone. This way the obtained extrusion will include all the points inside the original cone and has a simple definition of points inside and outside it. This fix fits well with the philosophy of the original algorithm where extrusions contain at least the closest points and only the minimum value is recorded.

\subsection{Surface curvature}

The signed distance function describes both positive and negative values. The extrusions must then be constructed on both sides of the surface features. We adopt the convention that the interior of the surface is negative and the exterior is positive. The outward extrusions are then along the feature normal and the inward extrusions in the negative normal direction. For all the triangle faces, prism extrusions extending in both positive and negative directions will encompass the area closest to that face. Work can be reduced, however, for the edge and vertex cases based on the curvature of the local surface. Mauch introduces the concepts of convex and concave features. 

\begin{figure}[h]
\centering
\begin{subfigure}[b]{0.25\textwidth}
\includegraphics[width=\textwidth]{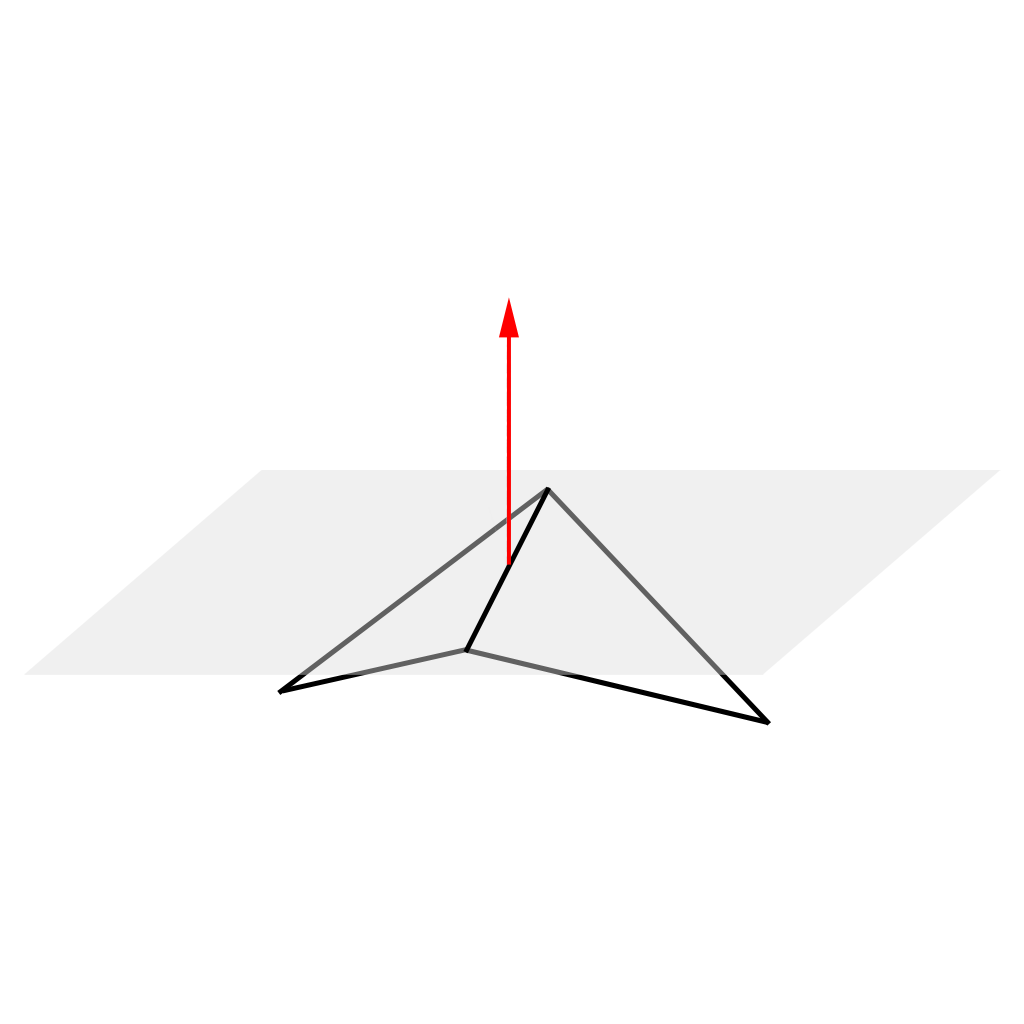}
\subcaption{convex edge}
\label{fig:curvature_a}
\end{subfigure}%
~\begin{subfigure}[b]{0.25\textwidth}
\includegraphics[width=\textwidth]{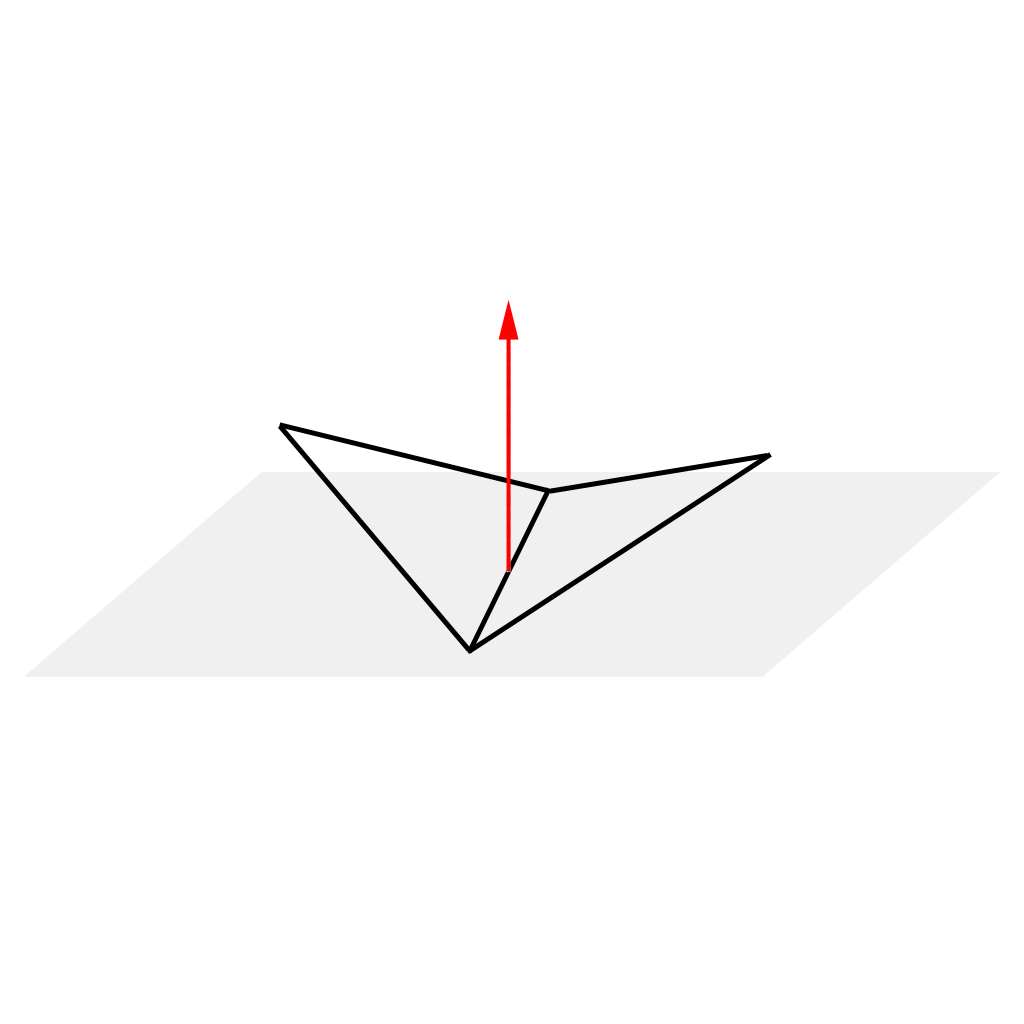}
\subcaption{concave edge}
\label{fig:curvature_b}
\end{subfigure}%
\begin{subfigure}[b]{0.23\textwidth}
\includegraphics[width=\textwidth]{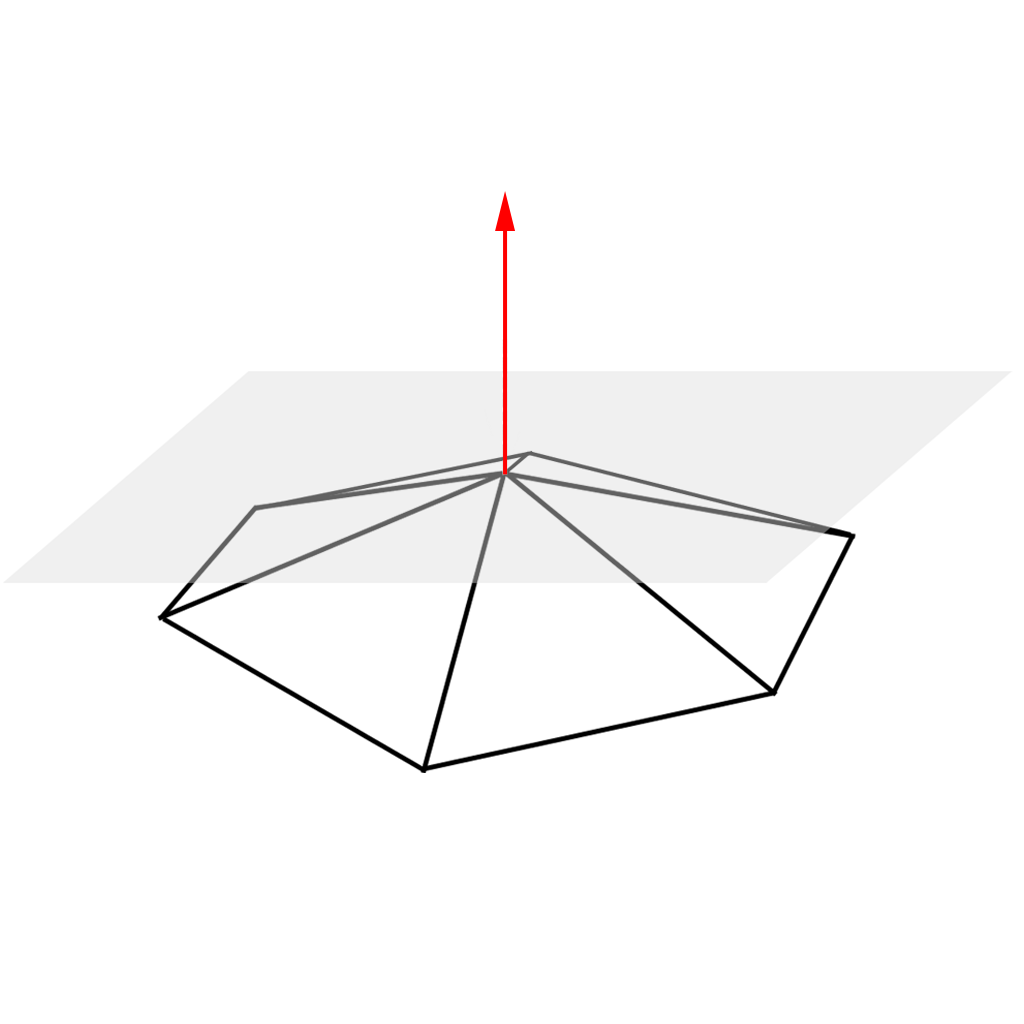}
\subcaption{convex vertex}
\label{fig:curvature_c}
\end{subfigure}%
~\begin{subfigure}[b]{0.25\textwidth}
\includegraphics[width=\textwidth]{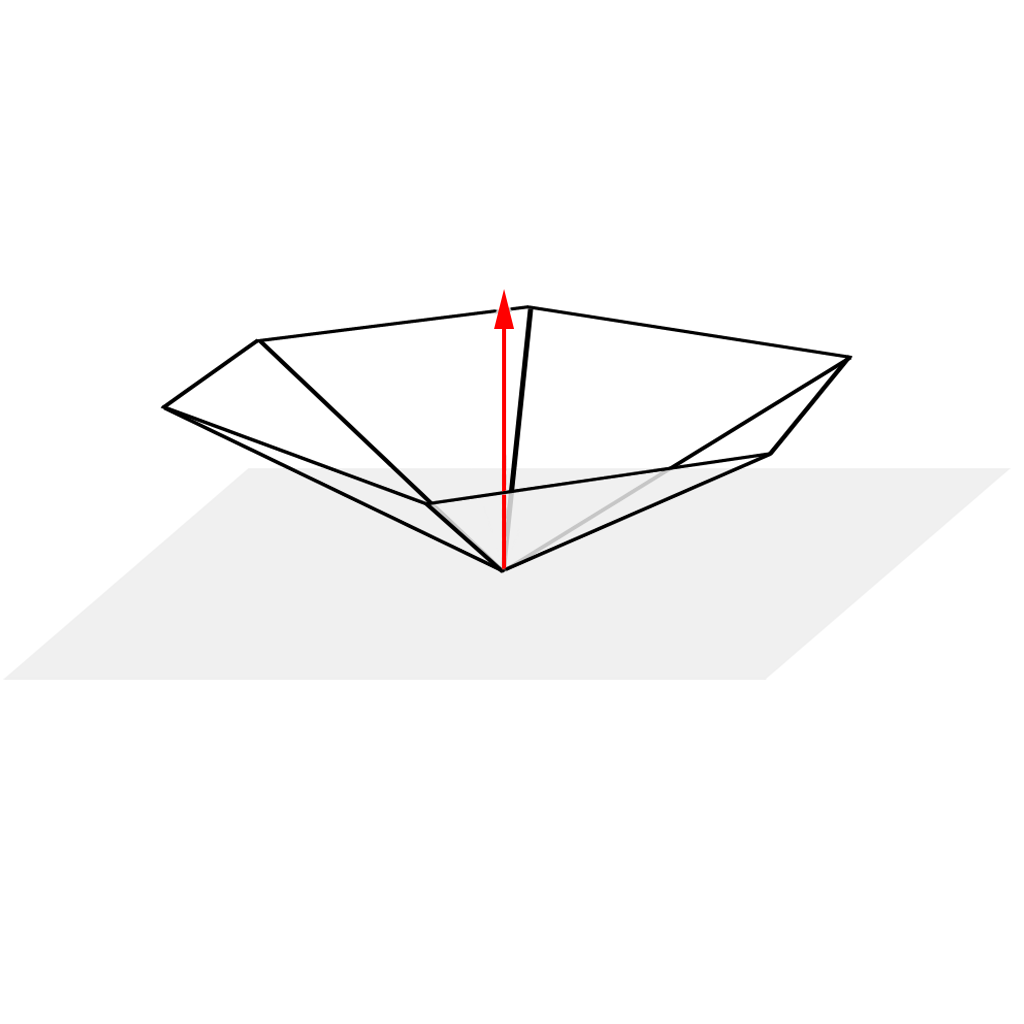}
\subcaption{concave vertex}
\label{fig:curvature_d}
\end{subfigure}%
\caption{The surface curvature is defined by the STL surface features and average normals. Convex edges and vertices have all of their neighbouring vertices below their average normal plane. Concave features have all of their neighbours above the plane.}
\label{fig:curvature}
\end{figure}

Taking the plane defined by the positive average normal of faces meeting at an edge and one of the edge vertices, an edge is convex if its endpoints lie above its neighbouring points, concave if they lie below and flat otherwise (figures \ref{fig:curvature_a} and \ref{fig:curvature_b}). The same is true for the coordinates of a vertex, which is convex when its neighbouring vertices all lie below the plane described by the vertex and the angle weighted average pseudonormal. A vertex is concave when its neighbours all lie above that plane, and flat  when they all lie on the same plane (figures \ref{fig:curvature_c} and \ref{fig:curvature_d}). Convex features will then only need positive extrusions, concave ones will only need negative ones and flat regions will need none as the surrounding extrusions from other features fill the area.

\subsection{Saddle}

A special case exists, however, where a vertex is neither convex, concave nor flat. These saddle points occur in common geometries and the original algorithm does not deal with the gaps left in-between the other extrusions, leading to regions of undefined distances and an incorrect signed distance field. A saddle point occurs when there are neighbouring vertices both above and below the plane described by the average pseudonormal of a vertex and the vertex itself as shown in figure \ref{fig:saddle}. A fix for this problem, as suggested by Peikert and Sigg\ \cite{peikert}, is to use both a positive and a negative extrusion at these vertices. These special cases will then warrant double the workload of other curvatures but we observe that this strategy fully covers the volume around the vertex in all of our test cases and leads to a consistent signed distance field around complex discontinuities. 

A question then arises about the shape of the extrusion from a saddle point vertex. While smooth convex/concave regions create a convex gap that is limited by the normals of adjacent faces, in saddle point cases this volume can be complex and difficult to assign an extrusion to. By using a cone defined by the pseudonormal as the axis and the most diverging normal on its side, the relative order and configuration of other normals does not matter and we have a well formed vertex extrusion. For saddle shapes our approach is to first generate the cone for the positive side and then reflect it in the negative pseudonormal direction for the interior distance generation.

\subsection{Ruff}

\begin{figure}[h]
\centering
\begin{subfigure}[b]{0.4\textwidth}
\includegraphics[width=\textwidth, trim={0 50 0 50},clip]{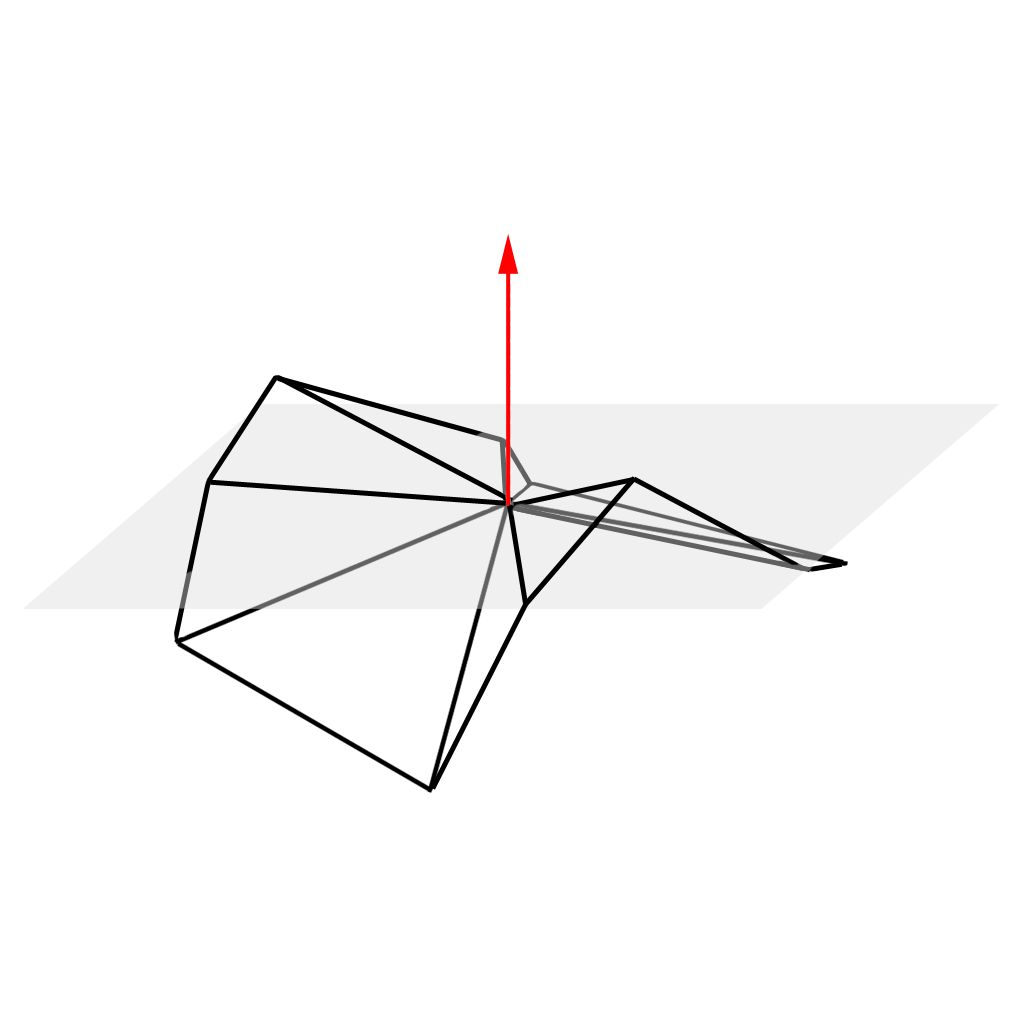}
\subcaption{saddle vertex}
\label{fig:saddle}
\end{subfigure}%
~\begin{subfigure}[b]{0.4\textwidth}
\includegraphics[trim = 0 60 0 50, clip, width=\textwidth]{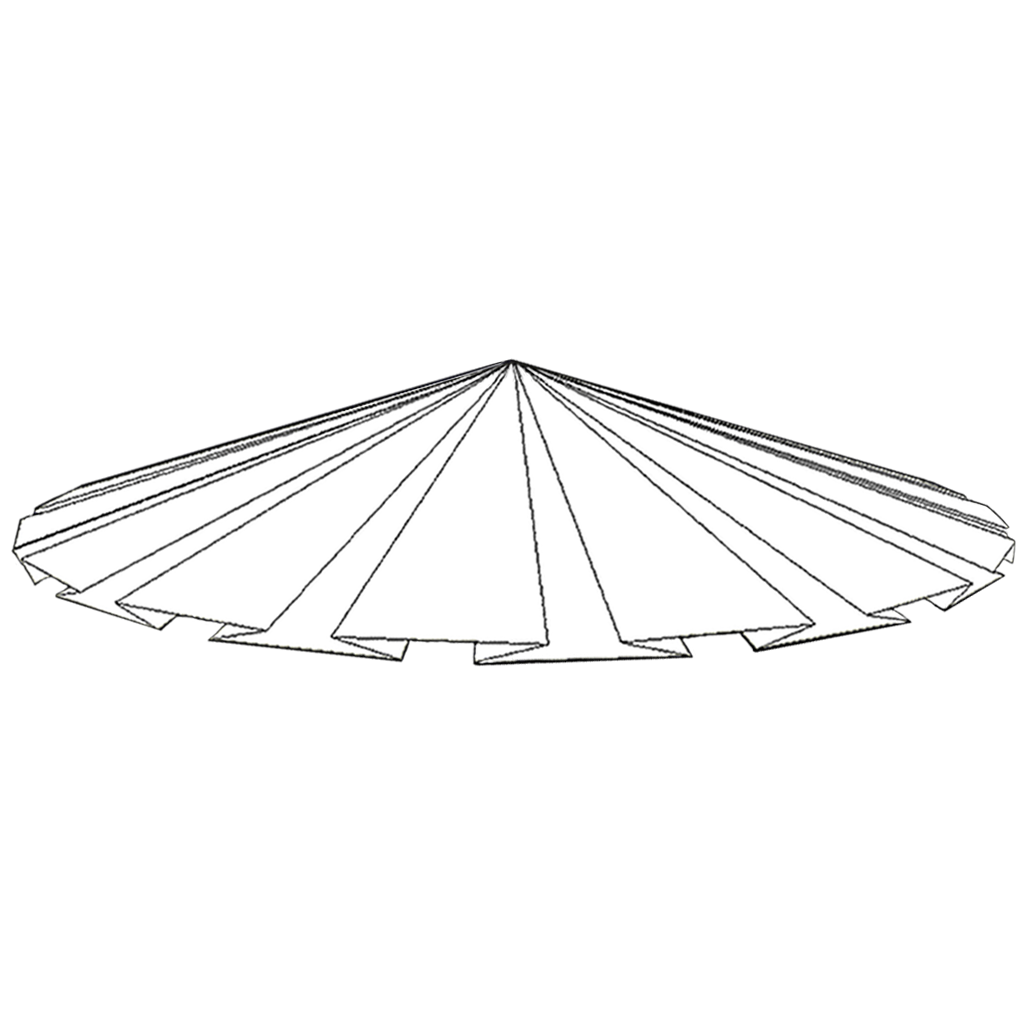}
\subcaption{ruff geometry}
\label{fig:ruff}
\end{subfigure}%
\caption{High curvature geometry. Saddle points are vertices where there are neighbouring vertices on both sides of the pseudonormal plane. Ruff geometries can be convex, concave or saddle vertices which feature normals that point to more than the half-space around the pseudonormal plane. These lead to complex gaps between extrusions from other features or call for extrusions that cause sign ambiguity.}
\label{fig:high_curvature}
\end{figure}

Even fully convex/concave vertices can have normals which do not define a simple region to extrude into. Consider the case shown in figure \ref{fig:ruff}. The illustrated ruff-like shape is a valid orientable triangulated surface where faces with almost opposite pointing normals meet at a single vertex. As all the neighbouring points end up being on the same side of the pseudonormal plane, the vertex is classified as convex. However, the space enclosed by the sum of the face normals extends below the pseudonormal plane at the vertex. Similarly, saddle points often, but not always, feature a collection of normals spanning more than the half-space. 

A simple solution for these cases is to only consider normals pointing to the same side of the pseudonormal plane as the average normal itself. The volume bounded by these positive pointing normals will be strictly less than the half-space above the pseudonormal plane which is coverable with a cone.  For our test cases this strategy fills the regions of ruff shapes and produces signed distance fields consistent with the input surfaces. It is possible, however, that a cone encompassing only positive pointing normals is not sufficient to cover the space between face and edge extrusions. In such cases a hemisphere can be constructed in the pseudonormal direction to cover the entire half-space above the vertex.

\section{Completeness of the CSC algorithm}
\label{sec4}

We show why the above mentioned procedures cover the immediate vicinity of the surface without leaving any gaps and why sign conflicts can be resolved unambiguously.

Consider generating the SDF for all of $\mathbb{R}^3$, so that there are no holes.  Space is divided by the surface into two regions, inside and outside, where moving from one region to the other along a continuous path  necessitates crossing the surface. For any point outside, the closest point on the surface to this point could lie on either a face, an edge or a vertex. 

\begin{figure}[h]
\centering
\includegraphics[width=0.25\textwidth]{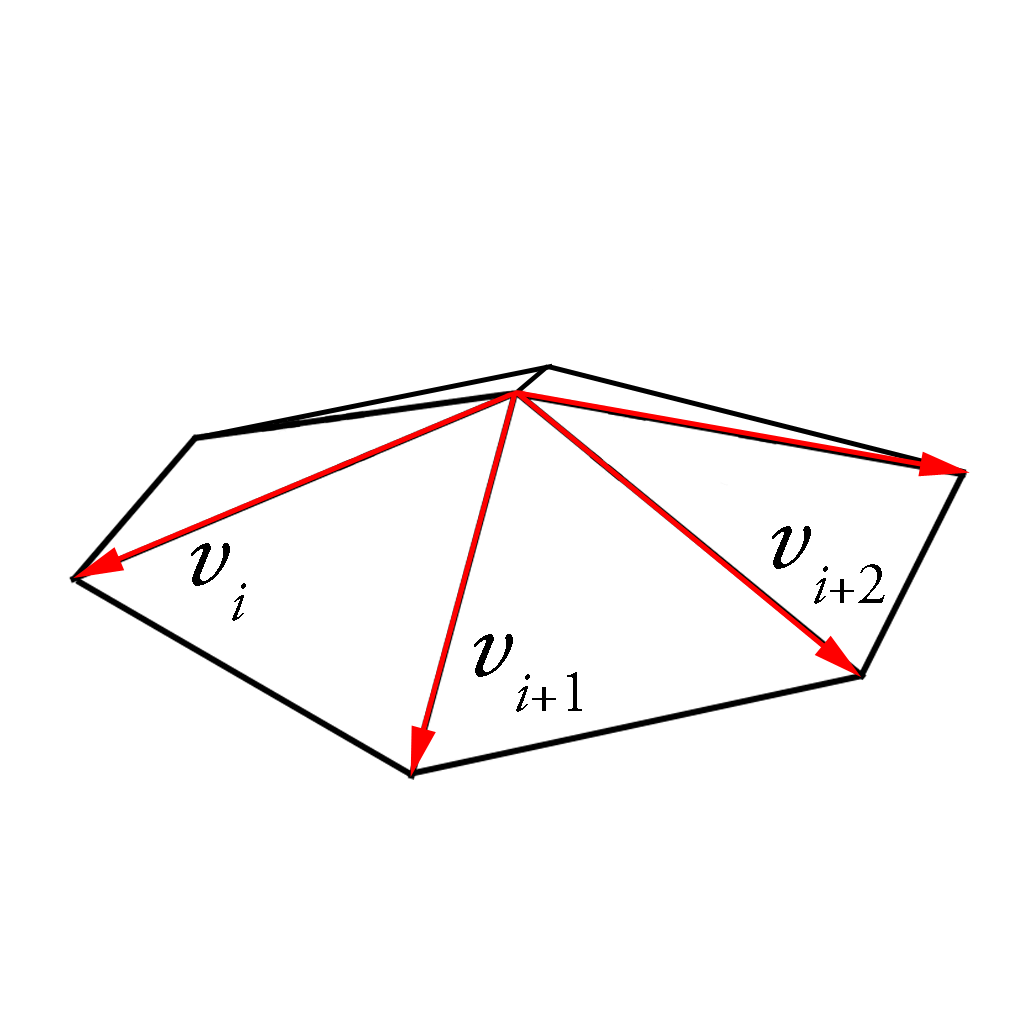}
\caption{Edge vectors at a vertex}
\label{fig:vertex_basis}
\end{figure}

Consider the simplified case of a single vertex, with edge vectors extending to infinity.  The vertex is at the origin, and the normalized edge vectors are labelled $\textbf{v}_1,\ldots,\textbf{v}_n$ (figure \ref{fig:vertex_basis}). We disallow faces of zero area (adjacent edge vectors that are parallel or antiparallel). A face extrusion from the $i^{\text{th}}$ face, is then the set of points given by

\begin{equation}
\lambda \textbf{v}_i + \mu \textbf{v}_{i+1} + \nu (\textbf{v}_i \times \textbf{v}_{i+1}),
\end{equation}

with $\lambda, \mu, \nu \ge 0$.  The neighbouring edge extrusion is given by

\begin{equation}
\lambda \textbf{v}_i + \mu (\textbf{v}_{i-1} \times \textbf{v}_i) + \nu (\textbf{v}_i \times \textbf{v}_{i+1}),
\end{equation}

with $\lambda, \mu, \nu \ge 0$.

The vertex extrusion is the positive spanning set of the surrounding face normals, namely, the set of points given by

\begin{equation}
\sum_i \lambda_i \textbf{n}_i,
\end{equation}

where $\textbf{n}_i = v_i \times \textbf{v}_{i+1} / |\textbf{v}_i \times \textbf{v}_{i+1}|$, 
$\textbf{n}_N = \textbf{v}_N \times \textbf{v}_1 / |\textbf{v}_N \times \textbf{v}_1|$, and $\lambda_i \ge 0$ for
each $i$.

\begin{minipage}{\textwidth}
A positive spanning set of vectors in $\mathbb{R}^3$ is either:
\begin{enumerate}
\item an infinite, convex pyramid, whose edges are the convex hull of the
  vectors,
\item an infinite wedge, when two of the vectors are antiparallel,
\item a half-space of $\mathbb{R}^3$,
\item the entirety of $\mathbb{R}^3$.
\end{enumerate}
\end{minipage}

A set of face normals can result in any one of these cases. The latter two cases can be obtained from ruff geometries.

For the SDF generation to be correct, these extrusions must fill the space to one side of the surface completely, since each extrusion is a superset of points where that edge, face or vertex is the closest point on the surface, and every point in space has at least one closest point to the surface, which must lie on \emph{some} feature.

From the above, it is clear that the procedure will not lead to any gaps: the edge extrusions are defined by the sides of the face extrusions, without any space between them, and the vertex extrusions are defined by the span of normals of the faces meeting at a vertex, the convex hull of which will always include all of the normals. There are no other features of a triangulated surface and in the absence of gaps in a closed surface, every point in its vicinity must exist in an extrusion. 

\begin{figure}[]
\centering
\begin{subfigure}[t]{0.24\textwidth}
\includegraphics[width=\textwidth]{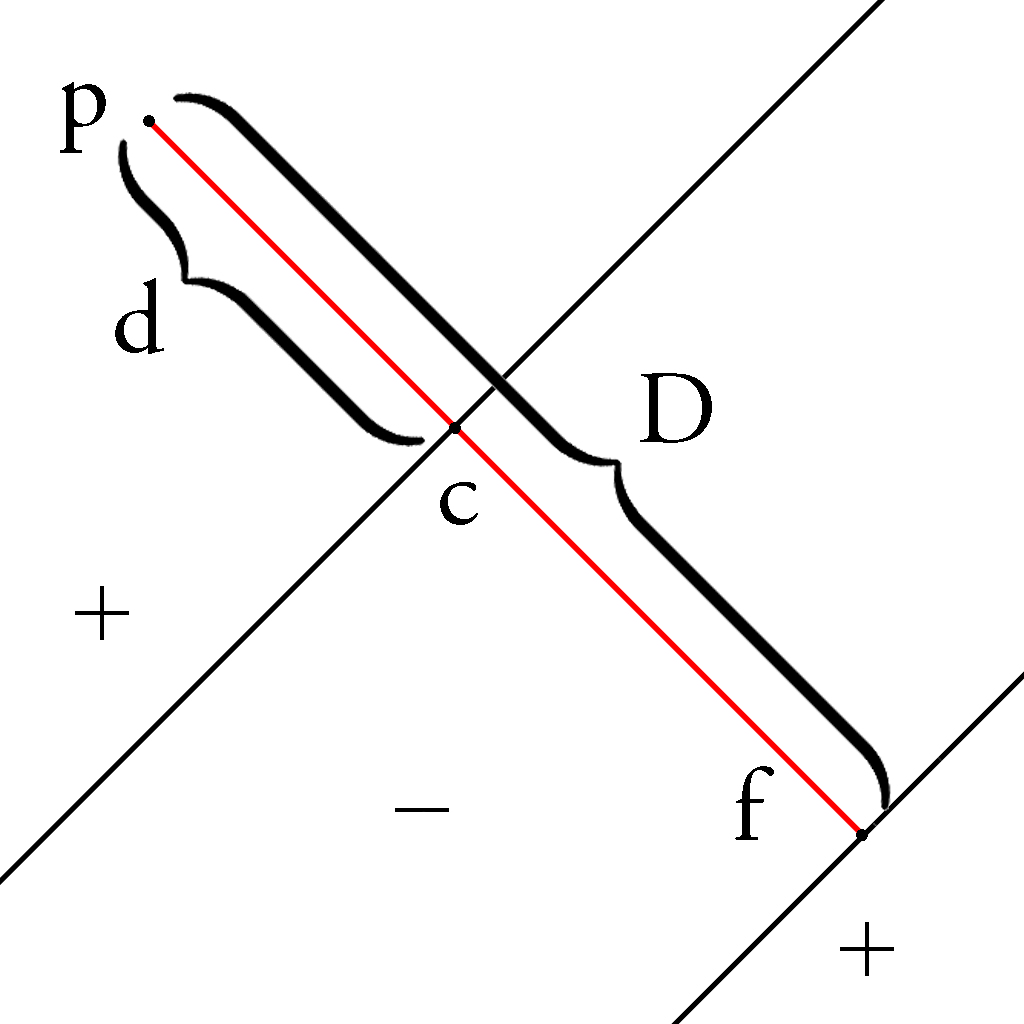}
\subcaption{incorrect sign at face when an extrusion from face f extends to point \textbf{p} where the ambiguity is resolved by the smaller distance d from a nearby face which the incorrect extrusion must cross at point \textbf{c}}
\label{fig:incorrect_face}
\end{subfigure}%
\,
~\begin{subfigure}[t]{0.24\textwidth}
\includegraphics[width=\textwidth]{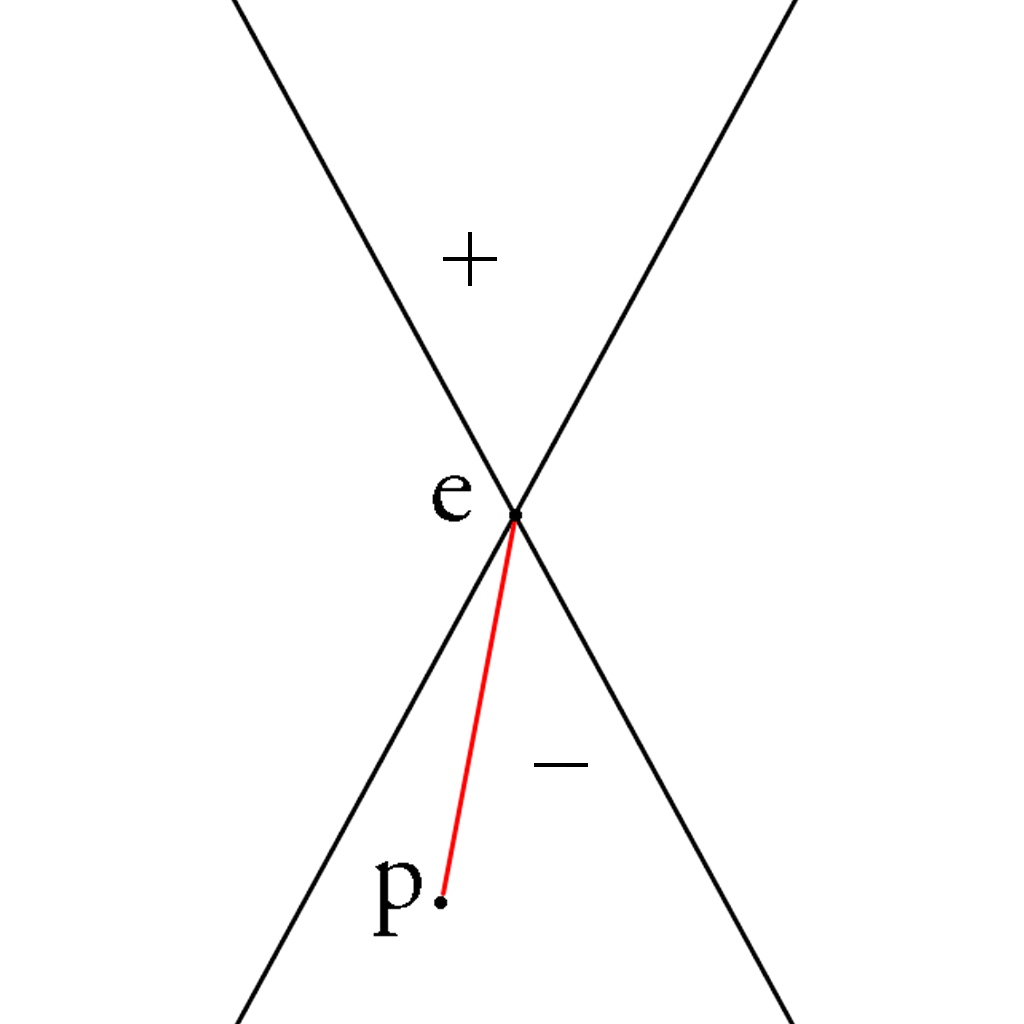}
\subcaption{incorrect sign at point \textbf{p} from the edge e will be overridden by another extrusion where the edge e cannot be the closest feature as its two extrusions of different signs are disjoint}
\label{fig:incorrect_edge}
\end{subfigure}%
\,
\begin{subfigure}[t]{0.24\textwidth}
\includegraphics[width=\textwidth]{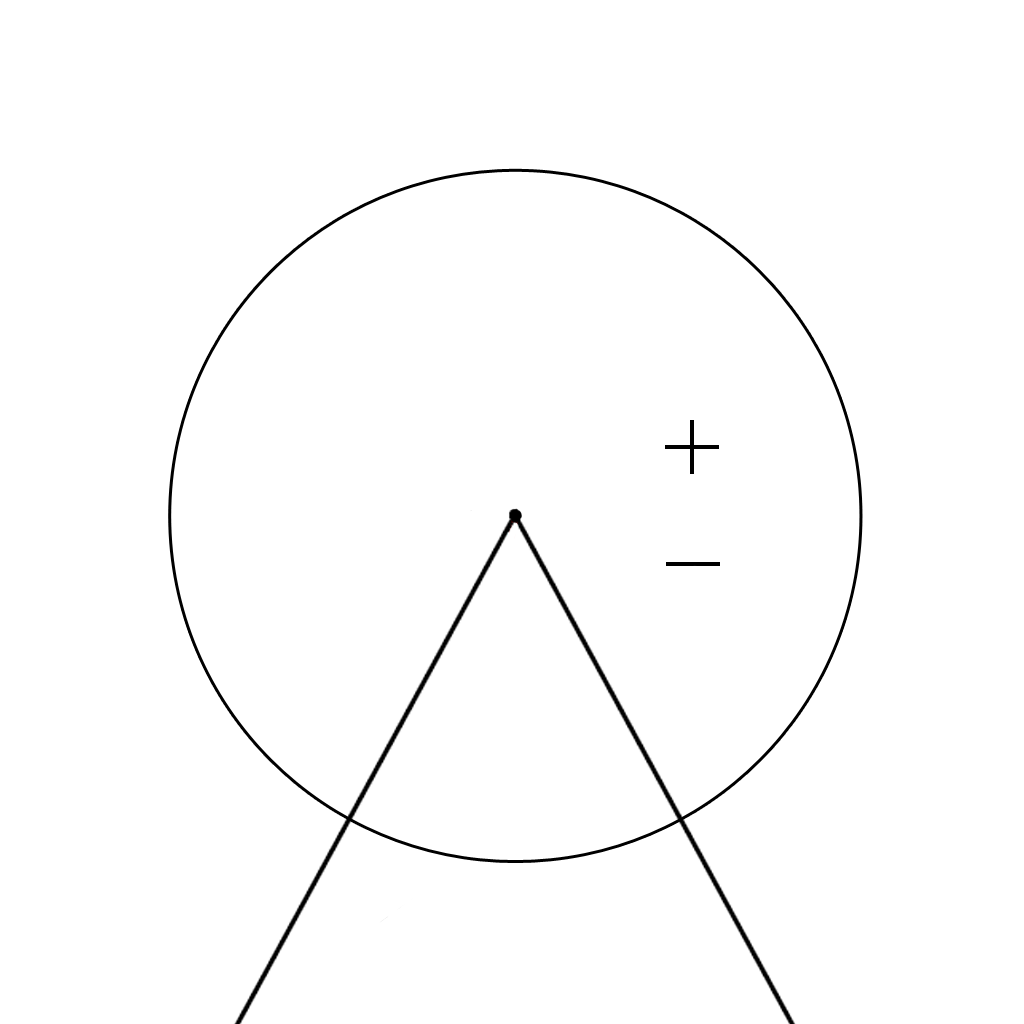}
\subcaption{ambiguous sign at vertex caused by normals spanning all of $\mathbb{R}^3$ which will lead to the positive and negative extrusions to overlap}
\label{fig:incorrect_vertex}
\end{subfigure}%
\,
\begin{subfigure}[t]{0.24\textwidth}
\includegraphics[width=\textwidth]{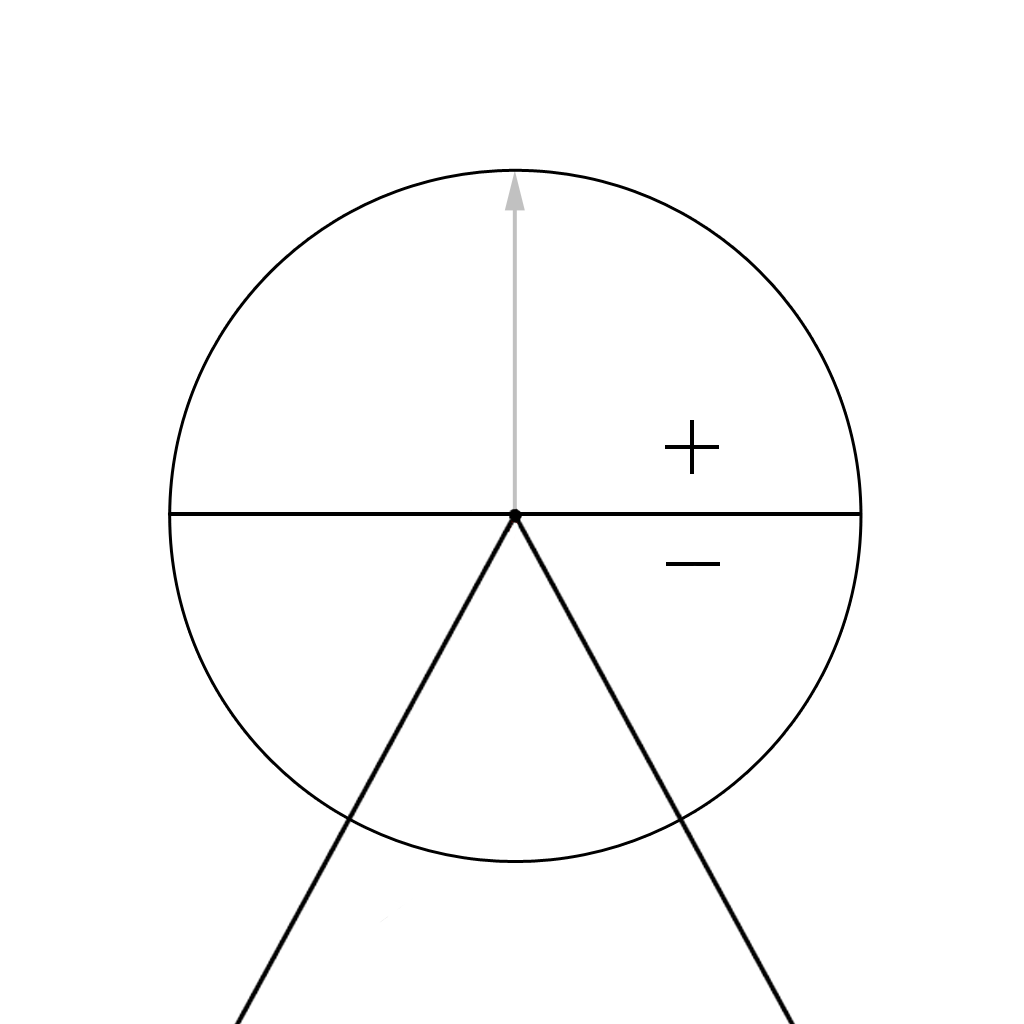}
\subcaption{assigning a hemisphere in the pseudonormal directions in such cases is enough to generate two disjoint volumes where the closest points inside extrusions are dealt with correctly}
\label{fig:hemisphere}
\end{subfigure}%
\caption{Sign ambiguity at features}
\label{fig:correctness}
\end{figure}

\subsection{Conflicts between positive and negative extrusions}

It is possible using the procedure described above for a given point of the domain (either inside or outside) to be both in a positive and a negative extrusion.  Only one of these can be of the correct sign, due to the orientability of the surface.  Choosing the extruded distance field with minimum absolute value at the point in question is enough to resolve the majority of these conflicts.  We describe each
case below.

\subsubsection{Incorrect distance information from face data}

Consider a point strictly on a face (not an edge or a vertex). As the surface normals $\textbf{n}_{\text{N}}$ point outside, there is some $\epsilon$ where $\epsilon \textbf{n}_i$ is within the exterior of the surface (i.e. there is always free space immediately adjacent to a face in the normal direction). Thus, if a sign conflict arises from a face extrusion at a point $\textbf{p}$, then a face extrusion must have crossed the surface. Consider figure \ref{fig:incorrect_face}: let the value of the incorrect extrusion from face f have absolute value D, and let the shortest distance from our point \textbf{p} to \textbf{c} where the extrusion crossed the surface be d, then $\text{D} > \text{d} + \epsilon$. This means that there is a closer point to \textbf{p} on the surface whose extrusion has the correct sign and the conflict does not cause any ambiguity.

\subsubsection{Incorrect distance information from edge data}

The situation is not as simple for edge vectors, since the face normals bounding an edge extrusion may point into the surface. However, note that if the edge extrusion with incorrect sign is given by

\begin{equation}
\{\lambda \textbf{v}_i + \mu (\textbf{v}_{i-1} \times \textbf{v}_i) + \nu (\textbf{v}_i \times \textbf{v}_{i+1})
\; : \; \lambda \ge 0, \; \mu, \nu > 0\}
\end{equation}
then the edge extrusion of the correct sign is
\begin{equation}
\{\lambda \textbf{v}_i - \mu (\textbf{v}_{i-1} \times \textbf{v}_i) - \nu (\textbf{v}_i \times \textbf{v}_{i+1})
\; : \; \lambda \ge 0, \; \mu, \nu > 0 \}.
\end{equation}

These sets are disjoint (figure \ref{fig:incorrect_edge}), meaning that if the incorrect edge extrusion conflicts with another extrusion of the opposite sign, the closest point on the surface cannot be on that edge at all, and there will exist an extrusion (from a face, vertex or another edge) with smaller absolute value. In other words, in cases where an edge extrusion would assign the wrong sign to a point, that edge cannot be the closest feature to that point, and in the absence of gaps, the point will also fall within some other extrusion with a smaller magnitude and the correct sign.

\subsubsection{Incorrect distance information from vertex data}

As with the edge data, it is possible for the face normals to point into the surface: that is, for $\epsilon \textbf{n}_i$ to be in the interior of the surface for all $\epsilon > 0$.

For the first three cases described above for vertex extrusions, they have the property that the corresponding extrusion of opposite sign is disjoint from the original one.  Similar to the case of the edge extrusion, this means that in the case of conflicting information due to the propagation of an incorrect sign from the vertex, there is a closer point from another extrusion of the correct sign.

For the final case where the face normals span the domain, there is a genuine ambiguity between the positive and negative extrusions: they are both propagating information with the same absolute value of the distance, but with conflicting signs (figure \ref{fig:incorrect_vertex}).

We solve the ambiguity by first computing an angle-weighted pseudonormal at the vertex. B\ae rentzen and Aan\ae s\ \cite{pseudonormal} show that this pseudonormal can be used as a discriminant for the surface at the vertex: if \textbf{p} is a point whose closest point on the surface is the vertex, then $\textbf{N}_\alpha \cdot \textbf{p} > 0$ when \textbf{p} is outside of the surface and $\textbf{N}_\alpha \cdot \textbf{p} < 0$ when \textbf{p} is inside the surface, where

\begin{equation}
\textbf{N}_\alpha = \sum_i \alpha_i \textbf{n}_i
\end{equation}

is the pseudonormal, and $\alpha_i$ is the angle of the face with normal $\textbf{n}_i$.

The extrusion is then performed only for the hemisphere oriented in the $\textbf{N}_\alpha$ direction for the positive extrusion, and in the $-\textbf{N}_\alpha$ direction for the negative extrusion (figure \ref{fig:hemisphere}). The positive and negative extrusions are disjoint, apart from the plane normal to $\textbf{N}_\alpha$. The closest point lying exactly on the plane can be excluded as if \textbf{p} is on the plane, then $\textbf{N}_\alpha\cdot \textbf{p} = 0$, and so by the discriminant property, \textbf{p} is on the surface and so is the vertex itself.

To show that this does not result in any gaps in the SDF, notice that a point outside of the positive hemisphere has $\textbf{N}_\alpha \cdot \textbf{p} < 0$, and so either \textbf{p} is closer to a point on the surface other than the vertex (and so must belong to another extrusion), or \textbf{p} is in the interior of the surface.

\subsection{Cone extrusion of vertices}

For cases where the normals at a vertex describe an infinite convex pyramid, we use a superset of the normals instead. The superset is formed by first constructing the angle-weighted pseudonormal $\textbf{N}_\alpha$ as described above, finding the face normal \textbf{n} at the vertex which diverges most from it as $i_{\text{min}} = \underset{i}{\text{arg\,min}} {|\textbf{n}_i \cdot \textbf{N}_\alpha|}$, where arg\,min gives the argument which minimises the result and constructing a cone with this normal lying on its side.

The pseudonormal is within the original convex pyramid, since it is a positive combination of the normal vectors. Since $\textbf{n}_{i_\text{min}}$ minimized the right-hand side of this expression among the $\textbf{n}_i$, the other $\textbf{n}_i$ are contained within it, as is the original positive span, since

\begin{equation}
\textbf{N}_\alpha \cdot \sum c_i \textbf{n}_i = \sum c_i \textbf{N}_\alpha \cdot \textbf{n}_i > \sum c_i
\textbf{N}_\alpha \cdot \textbf{n}_{i_\text{min}} = \textbf{N}_\alpha \cdot \textbf{n}_{i_\text{min}}
\end{equation}

for any positive coefficients $c_i$ with unit sum.

The above shows that the produced distance field will not have any gaps and that sign conflicts can be dealt with unambiguously. When limiting the algorithm to narrow bands, the same holds true as the extrusion distances are the same for all features.

\section{Scan conversion}
\label{sec5}

After generating the extrusions, we need to determine which domain cells lie inside them. This is a similar problem to scan conversion -- a method in computer graphics that transforms mathematically described polygons into rasterised shapes. Mauch describes how we can determine which discretely spaced cells are inside continuous extrusions in $3$D by reducing the scan conversion of a polyhedron to a series of $2$D problems where slices of the extrusions are scan converted on planes of mesh cells. 

However, the extrusions of the surface features are always either cones, hemispheres or convex prisms which can be rasterised in $3$D. For the prisms, we rasterise $3$D regions of the mesh based on the half plane test. This strategy is used to find out if a point is within a convex polyhedron by determining if it is on the same side of all of the polyhedron faces. We find that this change fits better with the overall strategy of multithreaded computation and gives rise to other optimisation techniques described in the implementation section. 

Let $\textbf{c}_{xyz}$ be a cell in the domain $D$ and let $E$ be an extrusion with inward pointing face normals $N$. Furthermore, let $\textbf{p}_{i}$ be a point on the $i$\textsuperscript{th} face of an extrusion. The half plane test to determine if a point is within a convex polyhedron can be written as:
\\

\begin{minipage}{\textwidth}
\begin{algorithmic}
\ForAll{$\textbf{c}_{xyz} \in D$} 
	\ForAll{$\textbf{n}_{i} \in N$} 
		\If {$(\textbf{n}_{i} \cdot \textbf{c}_{xyz} - \textbf{p}_{i}) < 0$}
			\State return false
		\EndIf
	\EndFor
	\State return true
\EndFor
\end{algorithmic}
\end{minipage}
\\
\\

In the implementation the scalar product is tested against some small value $\epsilon$, to take into account numerical errors and the spacing of Cartesian grid cell centres.

A point is within a hemisphere if it is within a specified distance of the sphere centre and on the correct side of a plane. Points inside a cone satisfy the condition:

\begin{equation}
\frac{\textbf{N}_{\alpha} \cdot \left( \textbf{p} - \textbf{v} \right) }{\left| \textbf{p} - \textbf{v} \right| } > \textbf{N}_{\alpha} \cdot \textbf{n}_{\text{md}}\quad,
\end{equation}

where \textbf{p} is the point, \textbf{v} is the vertex, $\textbf{N}_{\alpha}$ is the unit cone axis and $\textbf{n}_{\text{md}}$ is a unit vector on the side of the cone.

\section{Implementation}
\label{sec6}

In the following sections we outline the technological paradigm of GPUs and our implementation including data structures, geometry generation and the work scheduling and calculation of the signed distance field at discrete intervals. The code takes as input a stereolithogrpahy (STL) file and outputs a file that lists the SDF values in a $3$D grid. We are interested in generating a signed distance only at the immediate vicinity of the geometry and are then only concerned with the cells inside the union of feature extrusions extending to some small user-defined maximum distance from the surface.

The CUDA programming platform is a C-like interface by Nvidia for programming GPUs\ \cite{CUDA}. The SIMD architecture fits well with a Cartesian grid data structure with minimal dependence between different parts of an algorithm. CUDA allows the programmer to launch a large number of threads that are scheduled and executed on the graphics card. Though very powerful, GPUs require a strategy different to conventional CPU coding. The main speedup of CUDA comes from having a large bank of threads and fast content switching to mask resource fetching. 

\begin{figure}[h]
\centering
\includegraphics[width=0.45\textwidth]{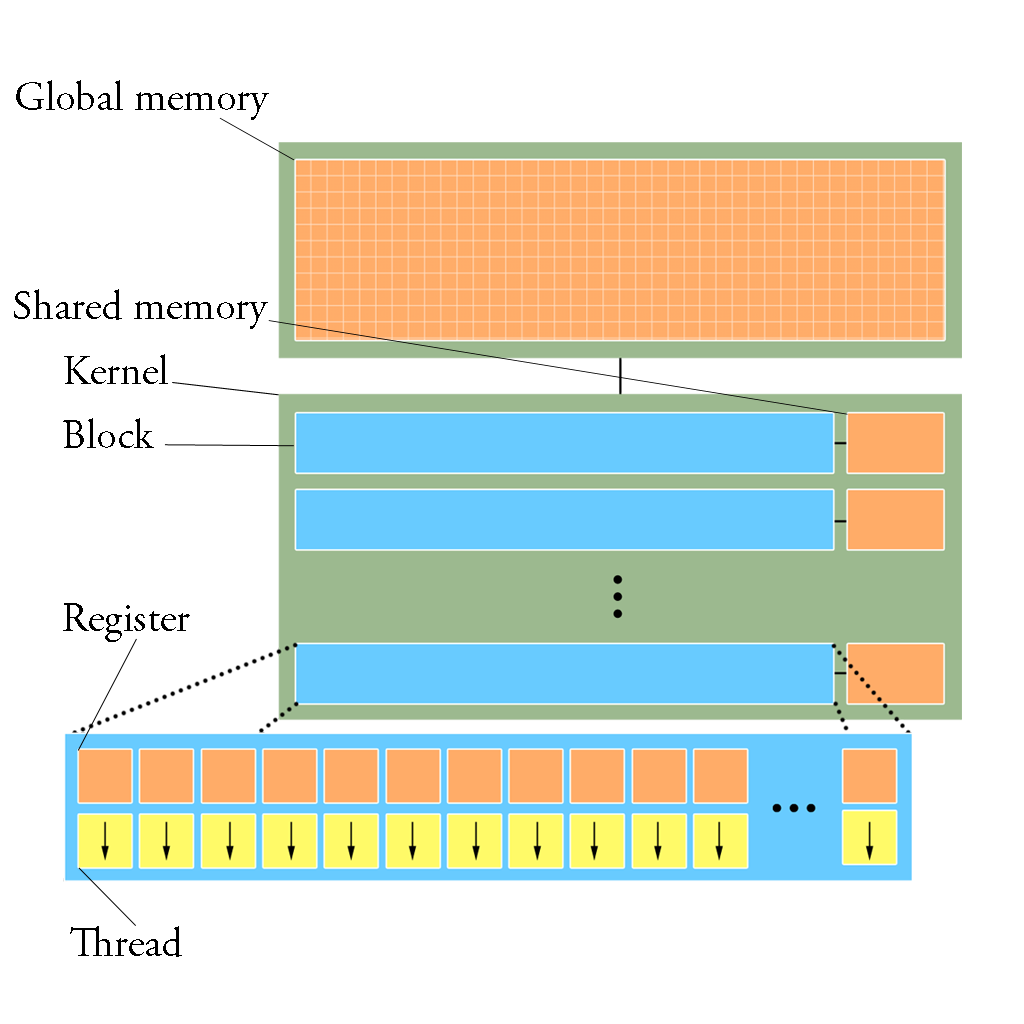}
\caption{Schematic of some CUDA concepts. The SDF domain will reside in global memory and kernels are launched to generate the signed distance values. A large number of threads grouped into blocks allow for high parallelism. There are limited banks of memory shared in blocks and smaller registers for each thread.}
\label{fig:CUDA}
\end{figure}

The programmer can launch blocks of threads that are handled and scheduled by the GPU and execute algorithmically simultaneously, but in practice, partially sequentially in an unspecified order. The threads are grouped into warps of $32$ which execute the same instruction simultaneously, and in the case of logical branching, parallelism may be lost. There is a bank of slow access global memory available to all threads, block specific shared memory and thread specific registers and limited scope caches. Figure \ref{fig:CUDA} shows a conceptual diagram of the different memory spaces and groupings from a programmer's point of view.

Memory access is very important to getting good performance, as the difference in bus speeds between the main global memory compared to registers and caches is many orders of magnitude. Conventional CUDA codes make use of programmer specified caching and read/write coalescence where units of threads access memory close together, thereby reducing the number of page transfer operations. The sparse layout of a surface in $3$D, however, gives rise to some interesting questions about memory access and work scheduling. This stems from data no-longer lying adjacent in logical groups in the domain, but along an arbitrary surface. It is not therefore immediately clear how to best address memory in a way that would minimise fetch transactions.

In the following sections, we will describe our approach to the original algorithm and the code that was produced. The implementation encompasses everything from reading in the STL file to outputting a result file. However, we only time the work done creating the internal geometry representation and the SDF generation. We will assume that the STL file describes a correct closed surface with no gaps between adjacent faces and no overlapping or flipped faces.

\section{Data structures}
\label{sec7}

The implementation starts with reading in the structured STL file that lists the vertices of each triangle face in a counterclockwise direction and an outward pointing normal: 

\begin{minipage}{\textwidth}
\begin{lstlisting}
facet normal ni nj nk
    outer loop
        vertex v1x v1y v1z
        vertex v2x v2y v2z
        vertex v3x v3y v3z
    endloop
endfacet
\end{lstlisting}
\end{minipage}

This information is used to construct a single entry in a Face object, three entries in an Edge object and three entries of a Vertex object. These objects are collections of the spatial coordinates of the vertices and normals of the features. We list them as structures of arrays where all the \textit{x} coordinates are followed by all the \textit{y} coordinates and finally the \textit{z} coordinates. We generate these objects on the CPU and copy them into the GPU global memory. While a Face object fully describes a triangle with a normal, Edge and Vertex objects need further processing to generate extrusions.

\subsection{Edge data}

An Edge object is created with two end-points of an edge and a normal of the triangle it was constructed from which is insufficient for an edge extrusion which needs two normals. Assuming a correct closed surface, there exists another entry with identical end points but a different normal. We would like to find matching pairs of edge features in the fastest possible way without checking each pair of endpoints against all the others. As the order of triangles in an STL file can be arbitrary, we would like to order the entries in the Edge object such that the pairs are next to each other. 
 
Sorting points in $3$D has no one correct solution, more so for pairs of points. One approach is to generate Morton codes for all of the points. Working with $32$ bit floating point values for all of the coordinate values, we can generate $30$ bit integer values called Morton codes for each $3$D point. These values will retain their relative position when sorted. Specifically, the sorted Morton codes will produce a Z-curve ordering. For our purposes, the actual order does not matter, only that identical edge features are positioned consecutively. 

An integer Morton code generated from the three floating point coordinate values of a vertex will designate its position in a $1$D array. We use a $30$ bit integer value stored in a $32$ bit \texttt{int} variable with the two highest bits set to $0$. The $32$ bit \texttt{float} coordinate values are first bit shifted to give $10$ digits preceding the decimal point. We then expand the three values using bitwise operations as shown in figure \ref{fig:expanding} or more concisely in code:
\clearpage

\begin{lstlisting}
x = (x | (x << 16)) & 0x030000FF;
x = (x | (x <<  8)) & 0x0300F00F;
x = (x | (x <<  4)) & 0x030C30C3;
x = (x | (x <<  2)) & 0x09249249;
\end{lstlisting}

\begin{figure}[h]
\centering
\includegraphics[width=0.8\textwidth]{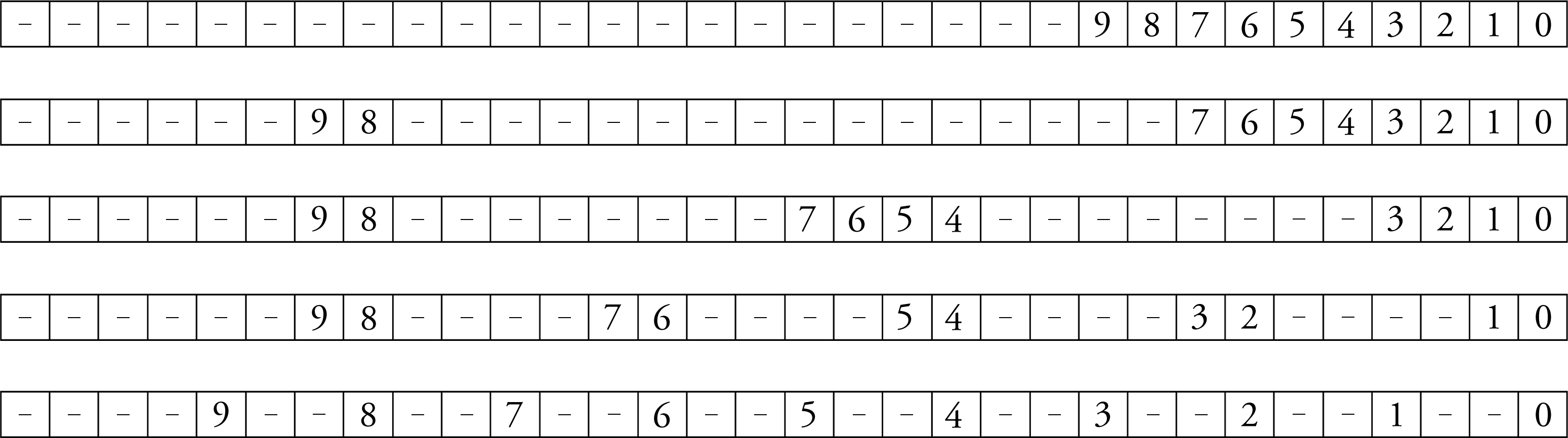}
\caption{Expanding a $10$ bit variable in four steps. The figure illustrates the movement of the original bit position values in the variable. (Adapted from\ \cite{morton})}
\label{fig:expanding}
\end{figure}

\begin{figure*}[h]
\centering
\includegraphics[width=0.8\textwidth]{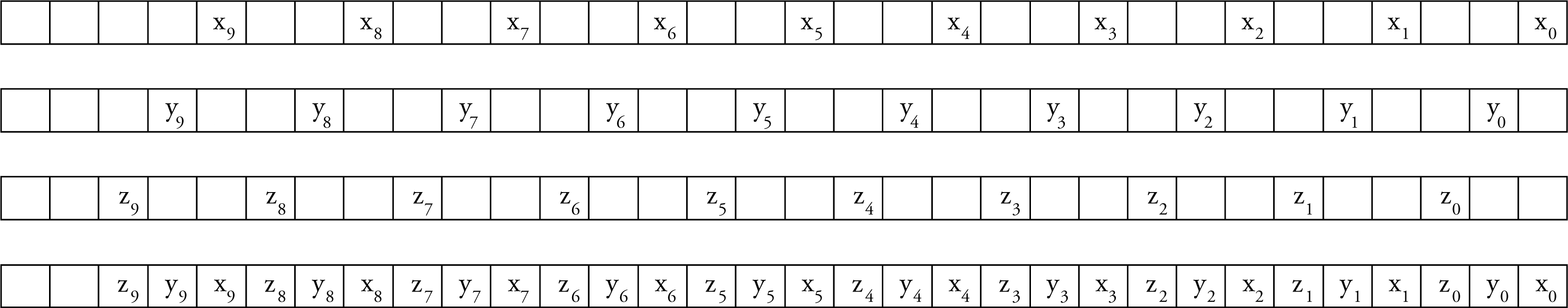}
\caption{Interleaving three bit-patterns into a single $32$ bit \texttt{int} variable using OR operations. (Adapted from\ \cite{morton})}
\label{fig:interleaving}
\end{figure*}

The resulting three \texttt{int} values are used to build the Morton code by shifting the \textit{y} and \textit{z} values further and interleaving all three into a single variable as shown in figure \ref{fig:interleaving}.

The expansion and interleaving of three $10$ bit values limits us to $1024^3$ unique values. We specify the Morton domain to encompass a space that is defined by the smallest \textit{x}, \textit{y} and \textit{z} Vertex values at one corner and the largest value Vertex at the opposite extreme. We launch a thread per Edge object and store the Morton codes in an \texttt{int} pointer on the GPU.

An Edge entry with two end points can then be transformed into a unique $60$ bit \texttt{long} value where two $30$ bit \texttt{int} values are concatenated such that the larger value takes up the high bit positions. Two edges with the same end points then have the same code and we can sort them to position identical edges next to each other. Because of the limited resolution of three $10$ bit values, this may still lead to a case where several Edge pairs have the same Morton code.
 
We use the Thrust\ \cite{thrust} library to sort a list of integer position indices based on the Morton code map using \texttt{thrust::sort\_by\_key}. We then reorder the Edge entries based on the indices using \texttt{thrust::gather}. This produces groups of edge entries with identical Morton codes being consecutive. Finally, we launch a thread for each group and sequentially traverse the collection with identical codes, reordering them if necessary such that identical edges are consecutive. In practice, this still allows for high parallelism but in theory, there can be a noticeable difference in the amount of work each thread does when the underlying geometry has large variation in the size of triangles.

\subsection{Vertex data}

The entries of the Vertex object are similarly incomplete. Each entry has data about the position of the vertex and the normal of the face it was generated from. In addition we know the angle between the two edges connecting at the vertex on that face. We also retain data about the two other vertices on the original triangle. We again employ the Morton code strategy from the Edge object. We generate $30$ bit \texttt{int} codes for each entry, sort a list of indices, reorder the Vertex objects and group identical entries together. This leads to an ordered list of entries where identical vertices are consecutive and each retains a unique normal and angle.

\section{Extrusion generation}
\label{sec8}

Once the data structures have been processed, we can generate the extrusions. There are three types of extrusions: prisms for the Face and Edge objects and cones or hemispheres for the Vertex objects. A prism is defined by six points and five sides. However, in order to tell if a point is within the area we are interested in, only four side normals and two points are needed (either two vertices on a face or the end points of a edge). A cone requires a point, an axis vector and the most diverging normal on its side. The hemisphere requires a point and a clipping plane.

\subsection{Face extrusion}

The prism from a face is constructed by first extruding the three corner vertices by the user specified distance in the normal direction. We then find side normals defined by the cross product of the counter-clockwise ordering of the vertices when viewed from the inside of the prism. These three normals and points on the sides can be used to describe the planes of the prism sides. We use two of the original face vertices as the points on the planes. We then save the smallest and largest coordinate values of the original and extruded vertices. This produces a cuboid axis aligned bounding volume (AABV) that contains the cell centres we wish to test for inclusion in the prism. The same is done for the negative extrusion of the original vertices in the flipped face normal direction. We end up with two Prism objects and their AABVs.

\subsection{Edge extrusion}

An edge extrusion is also a prism but extruded from a line. We start by determining if an edge is convex, concave or flat. Let the edge pair between vertices \textbf{a} and \textbf{b} be denoted by $\textbf{ab}$ and $\textbf{ba}$ with normals $\textbf{n}_1$ and $\textbf{n}_2$ respectively. We define a discriminant $d = (\mathbf{ab} \times \textbf{n}_1) \cdot \textbf{n}_2$. An edge is convex if $d > 0$, concave if $d < 0$ and flat if $d = 0$.

For convex edges we extrude the two end points of the edge by the specified distance in both the normal directions, thereby producing a prism. The four sides of the prism are described by the end points and the inward pointing normals constructed similarly to the face prisms. An AABV is also constructed to encompass the prism. For concave edges, the original normals are first flipped and the rest of the procedure is identical. For some geometries, it was discovered that a clipping plane at the base of the edge extrusion was needed to produce a smooth surface output. The plane is defined by the average normal of the edge and one of the end points.

\subsection{Vertex extrusion}

The regular vertex extrusion is a cone with a circle base. To construct it, we first scale the normals of the vertex entries by their angle. The collection of normals corresponding to a single vertex are then used to generate an average pseudonormal and find the largest angle between the average and the original normals.

All neighbouring vertices $\textbf{v}_N$ are tested to see if they are above or below the plane defined by the original vertex $\textbf{v}$ and the pseudonormal $\textbf{N}_{\alpha}$. Consider the discriminant $d = \mathbf{vv}_n \cdot \textbf{N}_{\alpha}$. Vertex $\textbf{v}$ is convex if $d > 0,\ \forall \textbf{v}_N$, concave if $d < 0,\ \forall \textbf{v}_N$, flat if $d = 0,\ \forall \textbf{v}_N$ and a saddle point otherwise.

For convex, concave and saddle shapes we store the vertex coordinates, the pseudonormal and the most diverging positive pointing normal. Similarly to prisms we define a bounding volume. The negative extrusion is constructed in the same way, but with a flipped average normal, where saddle points use a reflected positive extrusion. For ruff-like scenarios, we define an AABV of a hemisphere clipped at the pseudonormal plane.

When constructing the cone, the height is the user defined maximum distance. For sharp corners, it may happen that the cone base is very large and if positioned diagonally in the domain, would require a large AABV which would extend far beyond the region closest to the vertex. To avoid testing unnecessarily many cells, we take the intersection of the AABV of the cone and the bounding volume of a sphere with the radius of the maximum distance centred at the vertex. This leads to a smaller AABV and fewer cells to calculate the SDF for.

\section{Work scheduling}
\label{sec9}

After all of the extrusions have been generated, we come to the problem of how to schedule the SDF generation. For best performance, we would like to limit the number of calculations and memory transactions and do as much work as possible in parallel. The main variables in our software are the domain resolution, the desired maximum SDF distance and the number of surface features. Regardless of the extent of the computational domain, we only want to calculate the SDF for the sum of cells inside all of the extrusions, which often overlap. To determine intersection of a surface with the computational mesh, an SDF distance of around $5\Delta x$ is sufficient where $\Delta$x is the length of a cell in one dimension. To limit which cells check for inclusion in which extrusions, the code works only on the cells inside the bounding volumes. Work is therefore only done on the cells most likely to be within any extrusion and we limit the tested cell and extrusion pairs. There are two obvious approaches to parallelism in this case. 

The first is to check each cell in a bounding volume simultaneously. The start and end \textit{x}, \textit{y} and \textit{z} coordinates of the volume and the resolution of the domain are stored in the extrusion data. The number of cells the volume covers is then known and threads are launched according to the size of the volume and the domain coordinates of each thread can be determined from the limits of the bounding volume. All threads check if they are within the bounding volume's extrusion in parallel. For threads that are inside, the distance to the feature can be calculated, and threads with a smaller magnitude value than the previous one write their result to memory. This leads to warp divergence but as no action is taken for the other cases, there is no performance penalty. This implementation would launch a kernel per bounding volume where each thread works with the same data with the exception of their local coordinate data and the distance they calculate. For narrow band SDF generation of objects with uniform feature sizes, the bounding volumes are likely to be small and for high feature counts, the kernel launch will dominate the runtime, leading to poor scaling.

The second approach is to parallelise over the surface features. A thread is launched per extrusion and it loops through each cell location within the bounding volume, determining whether to write a distance value to memory. For narrow bands and high feature counts, the serial traversal of bounding volume cells is relatively lightweight and fast. However, many of the extrusions overlap and the implementation must ensure that the smallest magnitude value is found. For parallel computation, the writing must be atomic, which will introduce some serialisation when multiple threads are working with the same domain coordinates. We use the \texttt{atomicCAS} method to try to write a \texttt{float} value into memory if the recorded value at the address has a larger magnitude. This attempt continues until the local value is successfully written to memory or a smaller magnitude value is written by another thread. The effect of the serialisation depends on the input geometry and the thread scheduling but the impact on the runtime is small compared to the overall amount of work.

\subsection{Dynamic parallelism}

Consider, however, geometries with few features in high resolution domains (e.g. when simulating flow over a box). When the number of cells inside extrusions is significantly higher than the feature count, looping over cells inside bounding volumes dominates the runtime. While the overall generation time is usually on the order of seconds, there is still scope for improved performance by using a hybrid of the two approaches outlined above. Dynamic parallelism allows for kernels to be launched from the device. Wang and Yalamanchili\ \cite{wang} provide an analysis of CUDA dynamic parallelism. They show that there is potential for speedup in several problems with inhomogeneous workload but that the greater overhead of launching kernels on the device can negate the benefits. Tang et al.\ \cite{tang} discuss a dynamic platform which seeks to launch device side kernels only when the potential computation time outweighs the launch overhead. They show good speedup for several benchmark problems. A hybrid approach would then launch a single kernel from the host, assigning a single thread for each bounding volume which dynamically launch kernels with a thread per cell.

Launching kernels on the device has a greater overhead than host side launches but dynamic parallelism allows for more work to be done simultaneously. We therefore consider two alternatives: launching a thread per extrusion to loop through the cells or launching a thread per extrusion which will then itself dynamically launch a thread for each cell. The results section discusses the performance of both strategies.

\section{Calculating the signed distance field}
\label{sec10}

We allocate space in the GPU global memory for the $3$D domain as a row major $1$D \texttt{float} pointer. By storing the physical limits and width, height and depth information, we can find the \textit{x}, \textit{y} and \textit{z} coordinates of each cell from its offset in the pointer. 

The SDF calculation kernel first checks if a cell centre is within the extrusion in question by performing a half plane test against the sides of the polyhedron for prisms, or a discriminant test for cones and hemispheres. To avoid machine epsilon errors and issues with testing discrete grid positions against continuum planes, we compare the results against small values from $10^{-4}\Delta$x to $10^{-3}\Delta$x where $\Delta$x is the length of a cell in one dimension. If the point is within the extrusion, we calculate the distance to the feature. 

For a face with normal $\textbf{n}$ and a point $\textbf{p}$ on its surface, the distance to point $\textbf{c}$ can be found by $\textbf{n} \cdot \textbf{pc}$. If the absolute value is smaller than a user defined maximum, and if the previous magnitude at that cell centre is larger, we write the result to global memory with the appropriate sign depending on the extrusion. For edge extrusions, the distance is the distance to a line and for a vertex, it is the distance between two points in $3$D.

\begin{figure*}
\centering
\begin{subfigure}[t]{0.25\textwidth}
\includegraphics[width=\textwidth]{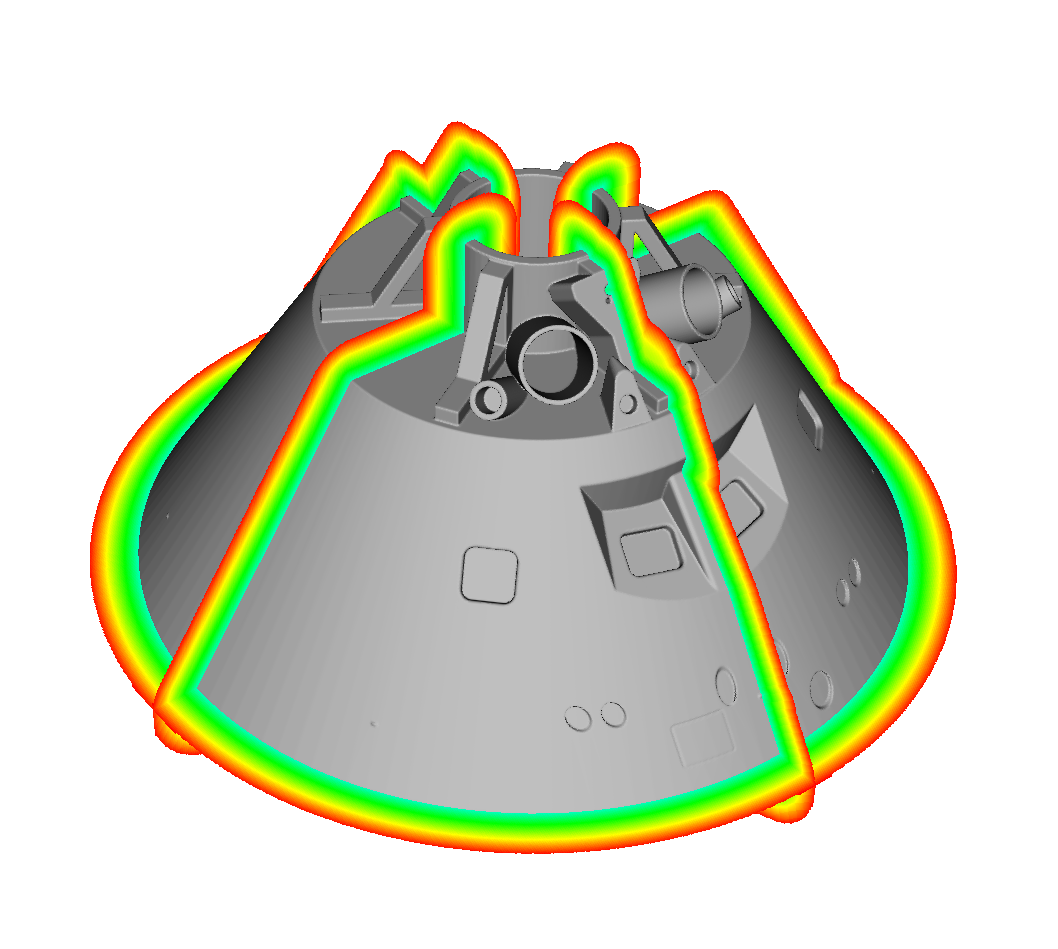}
\subcaption{Orion\ \cite{orion}}
\end{subfigure}%
\begin{subfigure}[t]{0.25\textwidth}
\includegraphics[width=\textwidth]{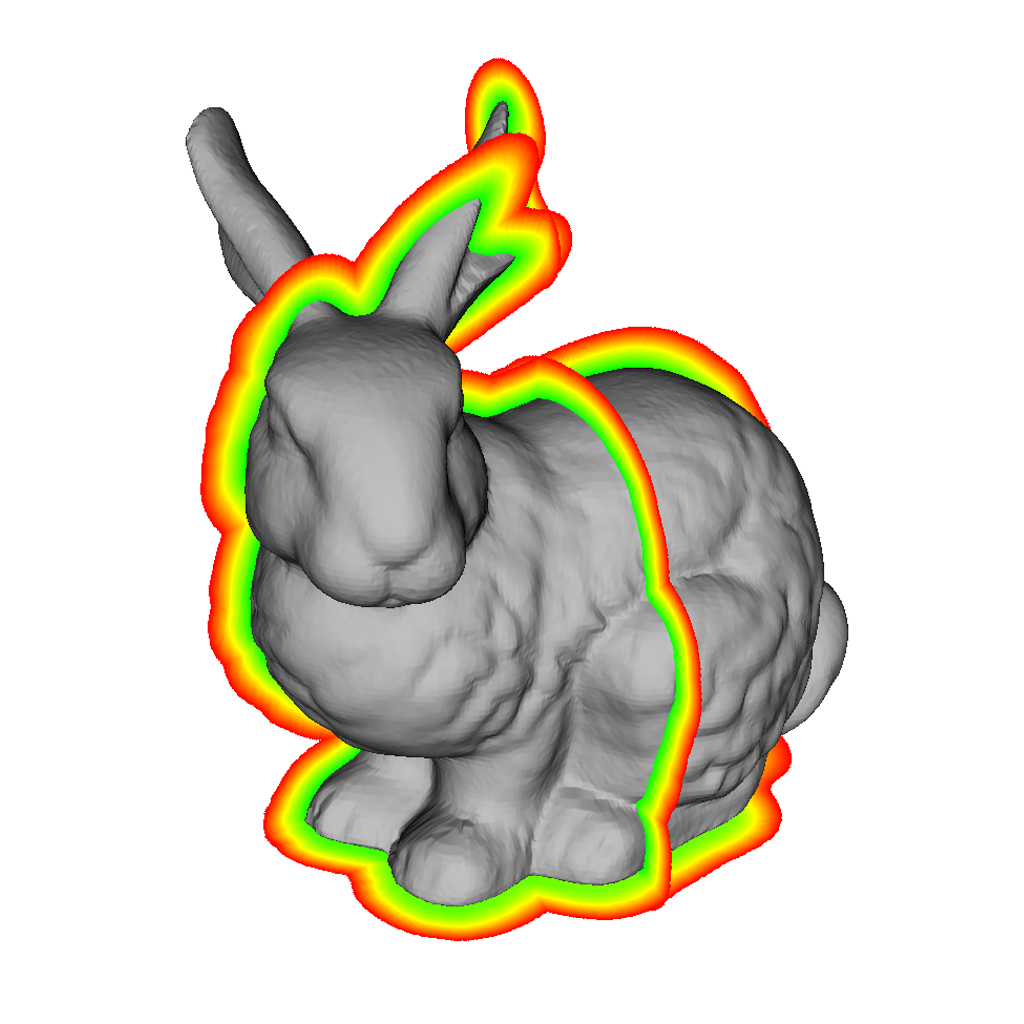}
\subcaption{Stanford Rabbit\ \cite{stanford}}
\end{subfigure}%
\begin{subfigure}[t]{0.25\textwidth}
\includegraphics[width=\textwidth]{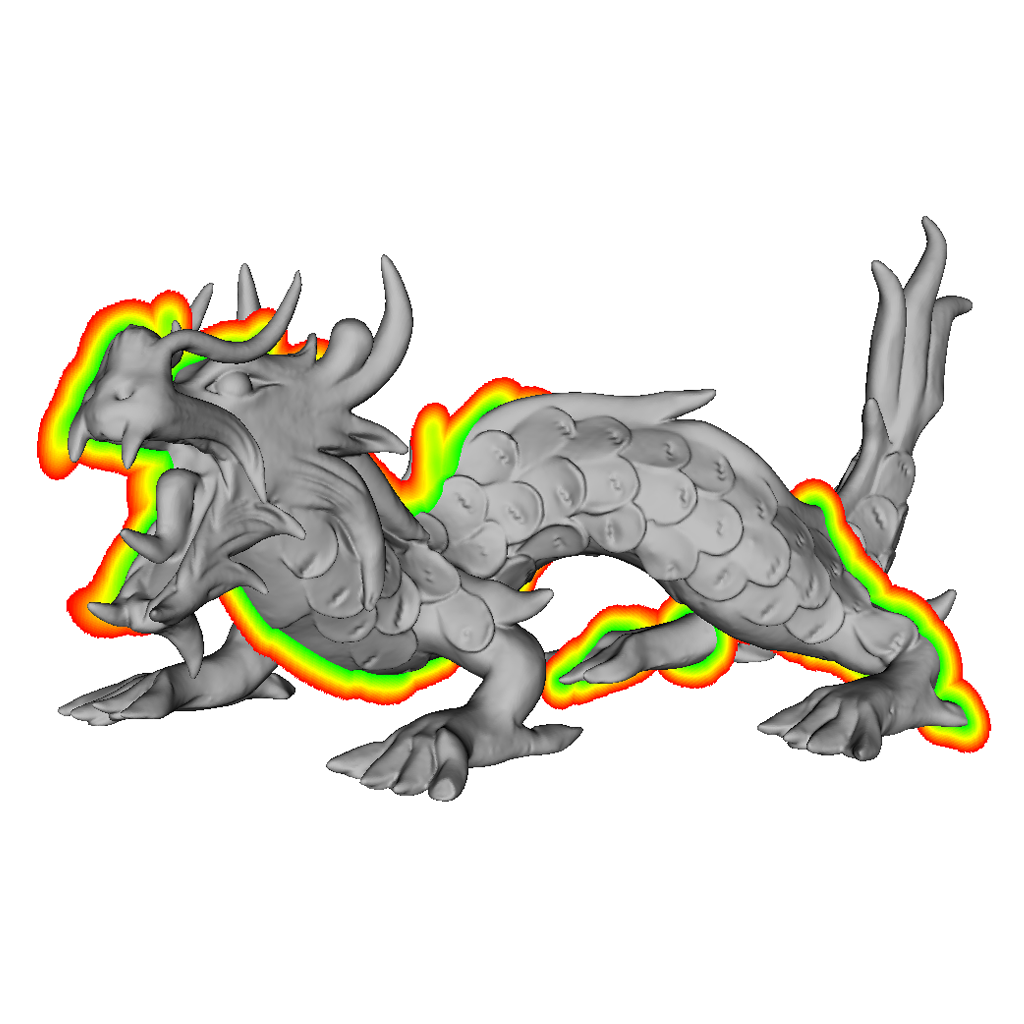}
\subcaption{XYZ RGB Dragon\ \cite{stanford}}
\end{subfigure}%
\begin{subfigure}[t]{0.25\textwidth}
\includegraphics[width=\textwidth]{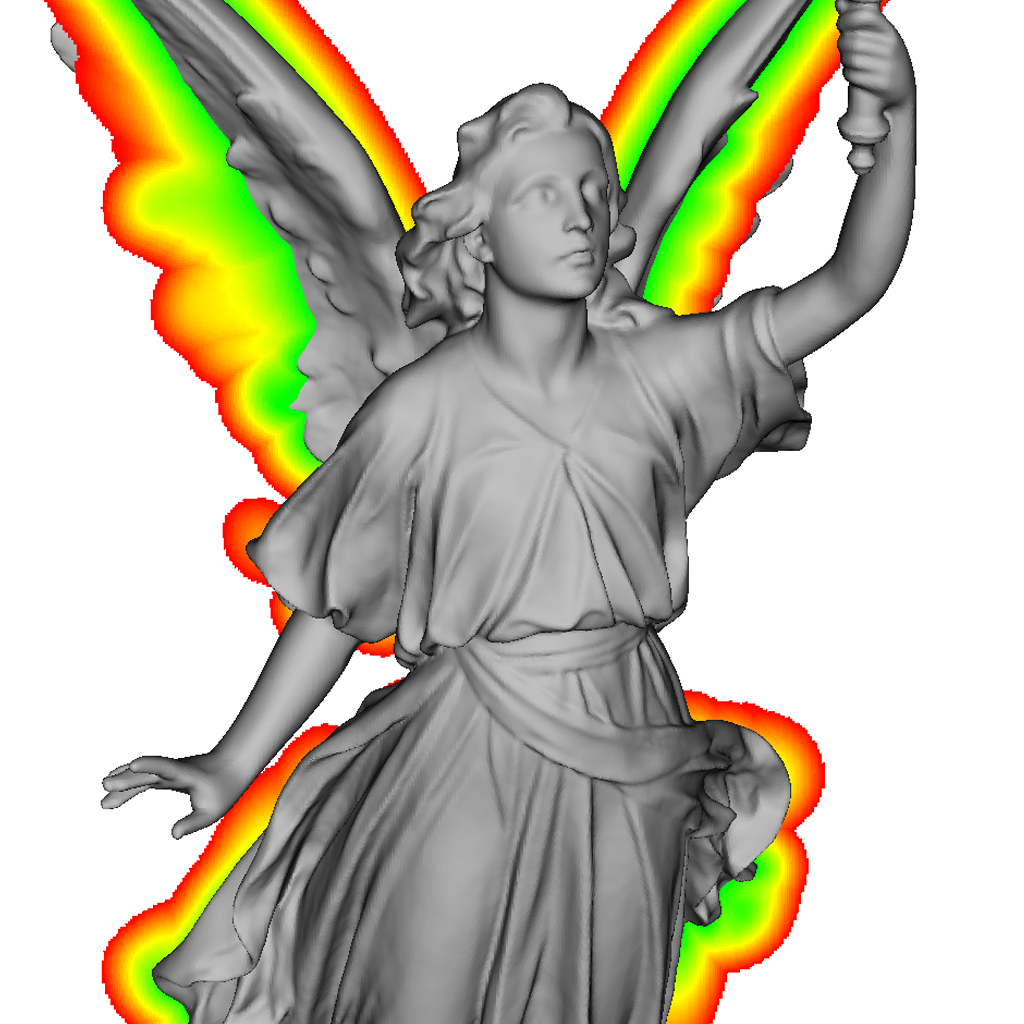}
\subcaption{Stanford Lucy\ \cite{stanford}}
\end{subfigure}%
\caption{Surface plots and SDF slices of test geometries. Narrow band signed distance fields were generated for complex shapes with varying feature counts on the GPU. The robustness and performance of the implementation allows for quick preprocessing times in various disciplines.}
\label{fig:examples}
\end{figure*}

\begin{figure}[h]
\centering
\begin{subfigure}[t]{0.4\textwidth}
\includegraphics[width=\textwidth]{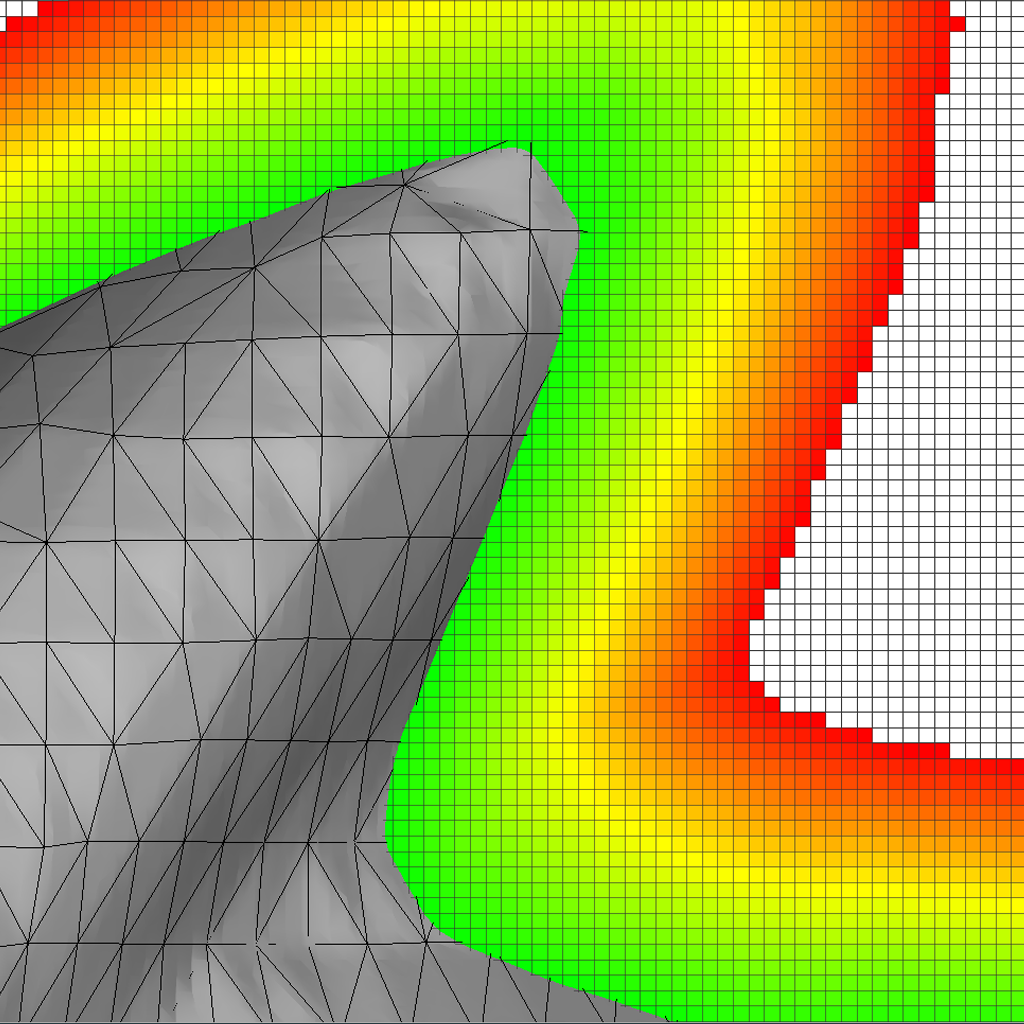}
\subcaption{level set (grey) with STL edges (black)}
\label{fig:resolution_1}
\end{subfigure}%
\,\,
\begin{subfigure}[t]{0.4\textwidth}
\includegraphics[width=\textwidth]{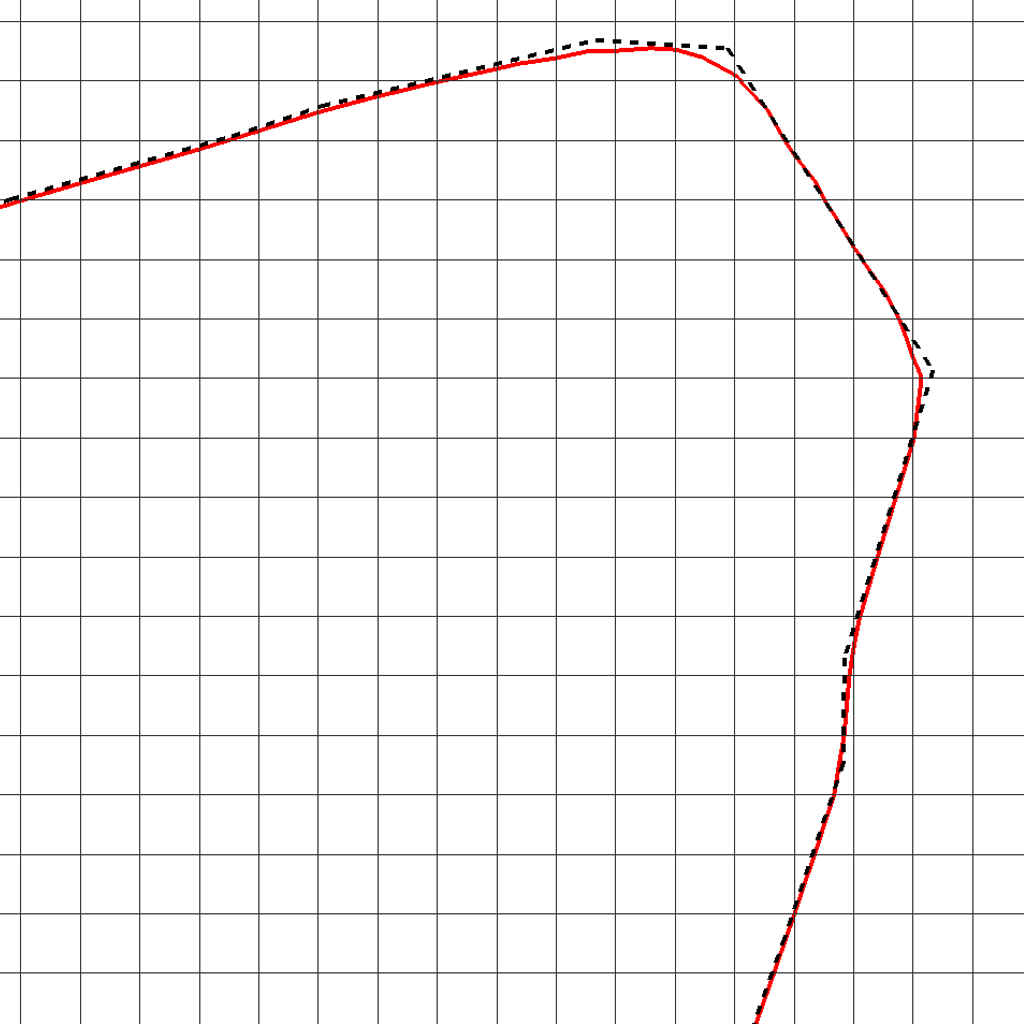}
\subcaption{level set (solid red) and STL slice (dashed black)}
\label{fig:resolution_2}
\end{subfigure}%
\caption{Results of Stanford rabbit ear at $\Delta x = 0.125$, $\text{distance} = 2$. (a) shows how the produced level set (grey) matches the mesh lines of the STL triangles (black). With sufficient domain resolution the code reproduces the sharp discontinuities of the input geometry. (b) shows how the slices of the level set (solid red line) and the STL (dashed black line) match to within $\Delta x$. Note that the visualisation software interpolates values and that the numerical accuracy is often higher than the image.}
\label{fig:resolution}
\end{figure}

\section{Results}
\label{sec11}

We present the results and timings of a number of test cases. The code was run on an Nvidia Tesla K$20$ card\ \cite{k20} with common STL geometries. We show surface plots with pseudocolour and isosurfaces of the SDF and list the preprocessing and distance generation times.

\subsection{Accuracy}

The produced code was validated against multiple common geometries which feature complex irregular surfaces as shown in figure \ref{fig:examples}. Figure \ref{fig:resolution} shows a zoomed-in region to illustrate the high resolution of the computational mesh, the SDF being set only in the immediate region of the surface (\ref{fig:resolution_1}) and how the produced surface matches the input mesh with an expected error of the order of the cell size (\ref{fig:resolution_2}). (The visualisation software interpolates both the SDF values and the surface slices, which makes the image a close approximation, not an exact reproduction.)

\begin{figure*}[h]
\centering
\begin{subfigure}[t]{0.4\textwidth}
\includegraphics[width=\textwidth]{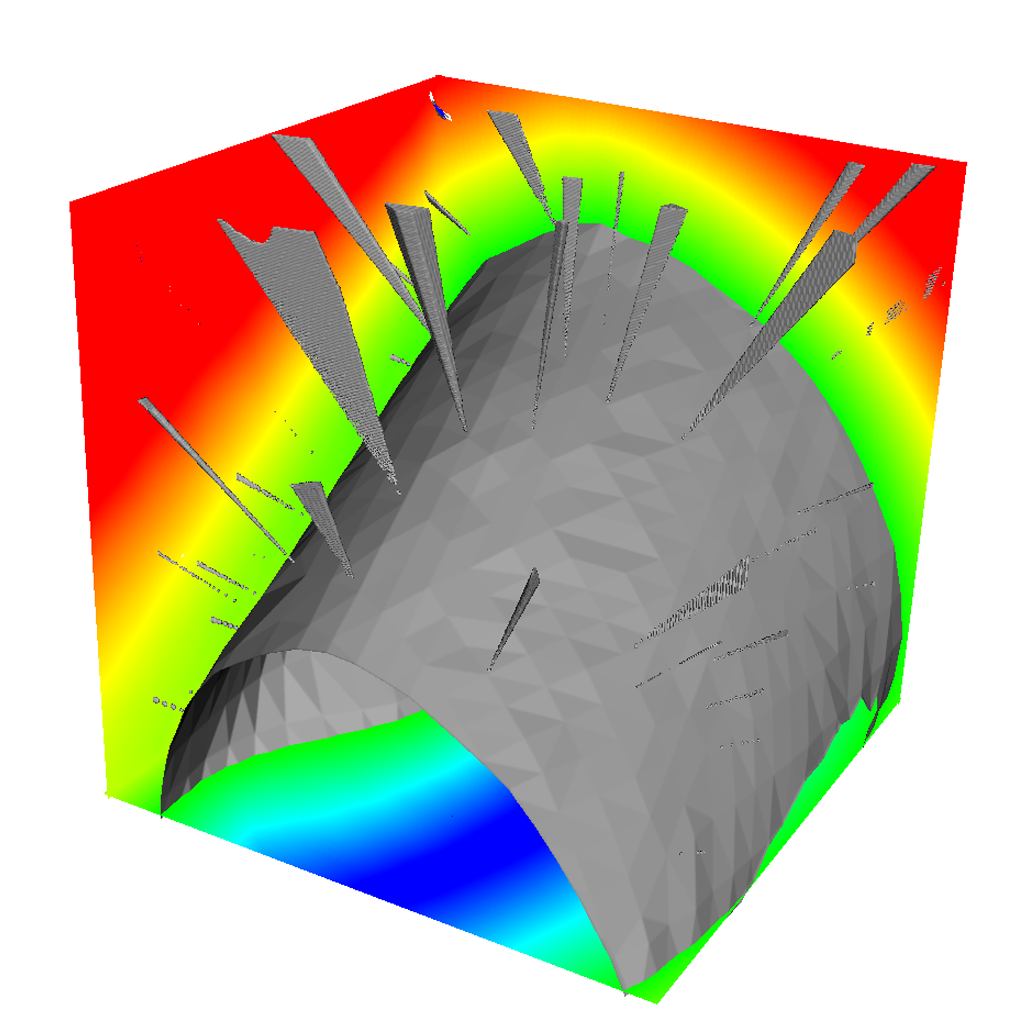}
\subcaption{Errors when only considering convex and concave vertices}
\label{fig:ear_high_wrong}
\end{subfigure}%
\,
\begin{subfigure}[t]{0.4\textwidth}
\includegraphics[width=\textwidth]{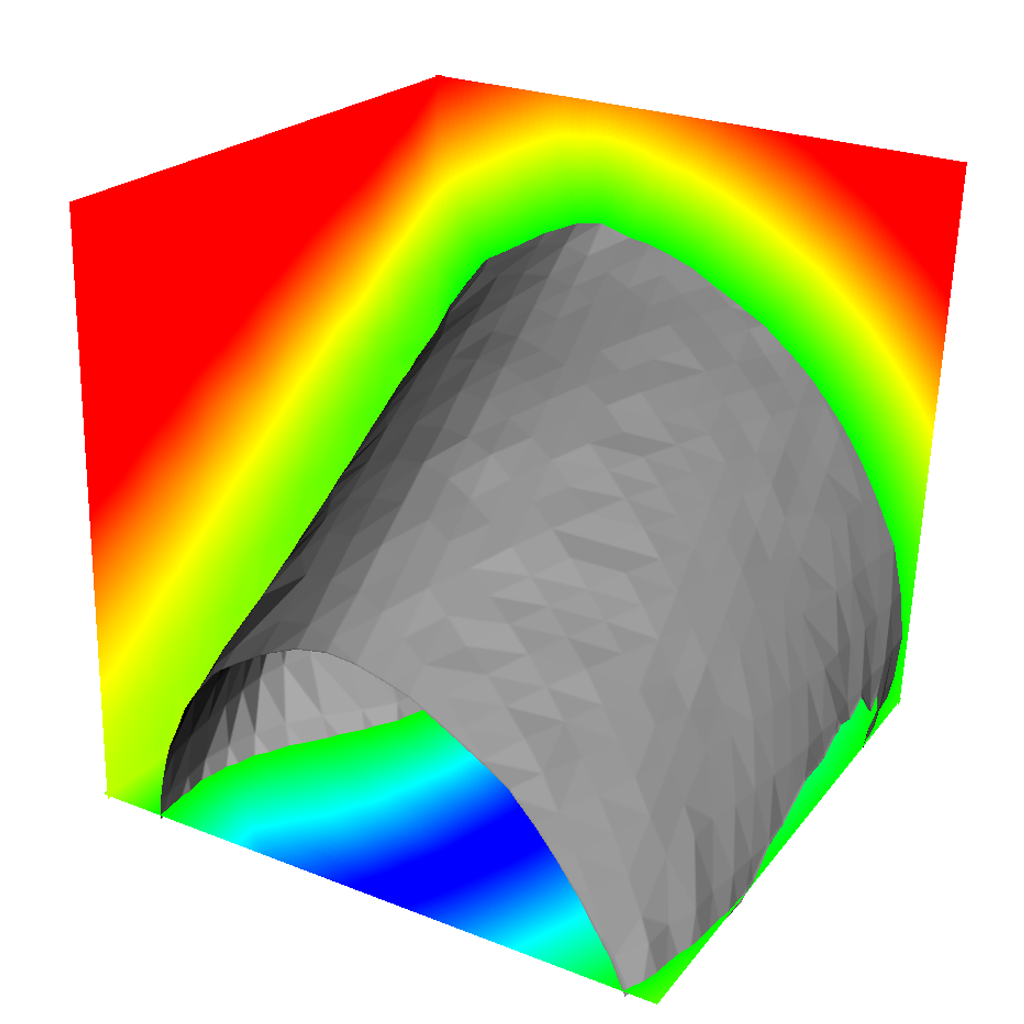}
\subcaption{Correct result when addressing saddle points}
\label{fig:ear_high_correct}
\end{subfigure}%

\caption{Results of the Stanford rabbit ear at  $\Delta \text{x} = 0.03$ show the issues with saddle points. When only assigning extrusions to convex and concave vertices, holes are left at saddle points. As no extrusion assigns a correct distance or sign, other extrusions can bleed into these regions. This can result in pyramid artifacts protruding from the surface where negative extrusions are never overwritten (a). At lower resolutions the errors can appear as artifacts farther away as the region closer to the surface gets the correct sign from nearby extrusions, but the ends of the interior extrusions are left uncorrected. (b) shows the correct SDF when assigning extrusions to saddle points.}
\label{fig:ear}
\end{figure*}

Figure \ref{fig:ear_high_wrong} illustrates the errors produced by only considering convex and concave vertices at the right ear of the Stanford rabbit geometry\ \cite{stanford}. When not addressing saddle points, gaps are left into which nearby extrusions may extend. As these values are never overwritten, artifacts may be produced. When a negative extrusion is not overwritten by a smaller magnitude positive extrusion on the outside of the surface, pyramid like protrusions are created in the level set. These errors may also appear as farther away spheres when the values near the surface are covered by neighbouring positive extrusions. Figure \ref{fig:ear_high_correct} shows the correctly produced SDF by generating extrusions on both sides of saddle vertices by building a cone around positive pointing normals and reflecting them to the negative pseudonormal direction. 

\begin{figure*}
\centering
\includegraphics[width=0.45\textwidth]{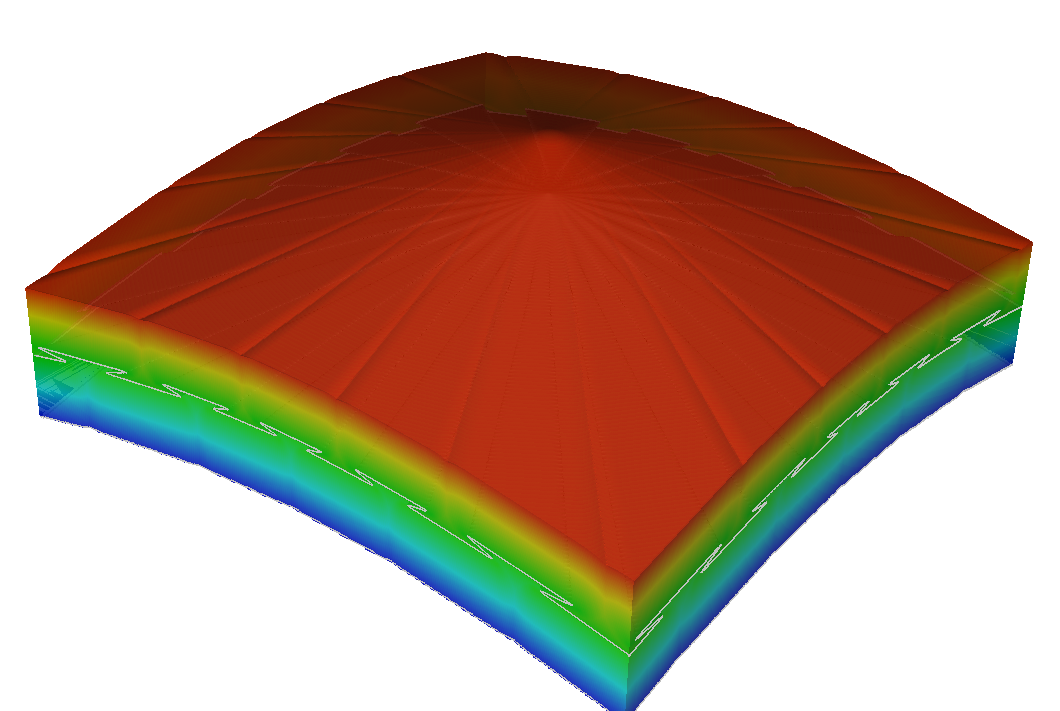}
\caption{A continuous SDF around a ruff geometry. A ruff vertex is classified as convex but the normals of the faces meeting at it span $\mathbb{R}^3$. By considering only the normals which point to the positive side of the pseudonormal plane, a strictly less than half-space volume can be filled with the distance to the vertex. The result is a continuous signed distance field around the surface.}
\label{fig:ruff_fix}
\end{figure*}

\begin{figure*}
\centering
\begin{subfigure}[t]{0.25\textwidth}
\includegraphics[width=\textwidth]{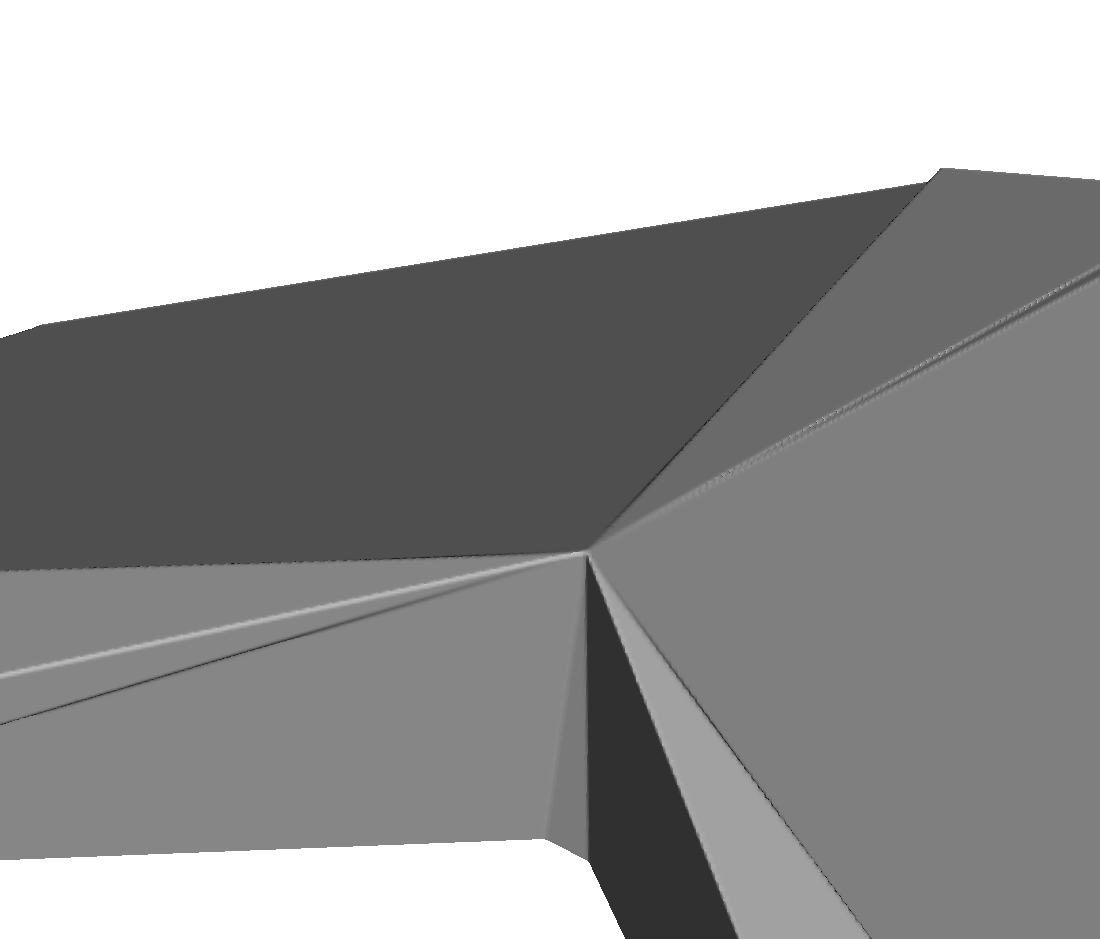}
\subcaption{Pathological surface}
\label{fig:pathological_1}
\end{subfigure}%
\begin{subfigure}[t]{0.25\textwidth}
\includegraphics[width=\textwidth]{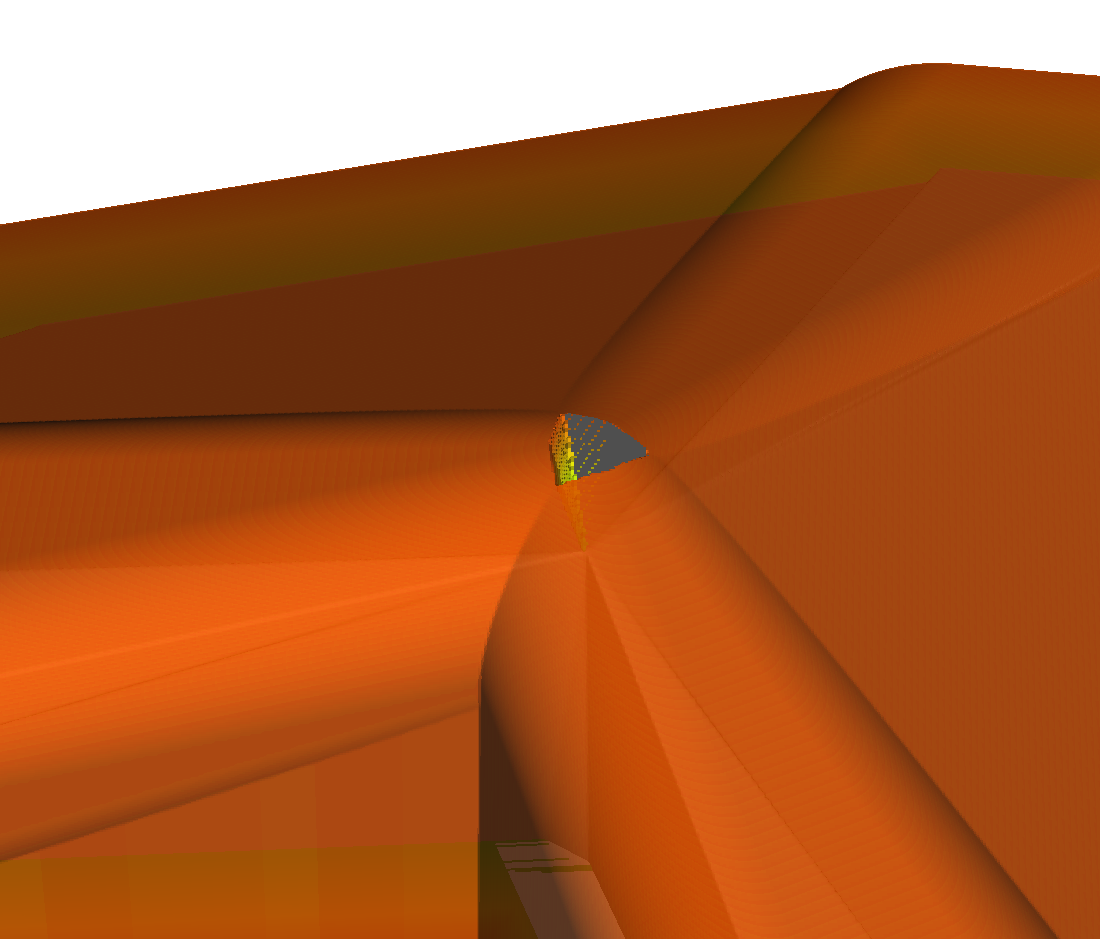}
\subcaption{Hole left at vertex}
\label{fig:pathological_2}
\end{subfigure}%
\begin{subfigure}[t]{0.25\textwidth}
\includegraphics[width=\textwidth]{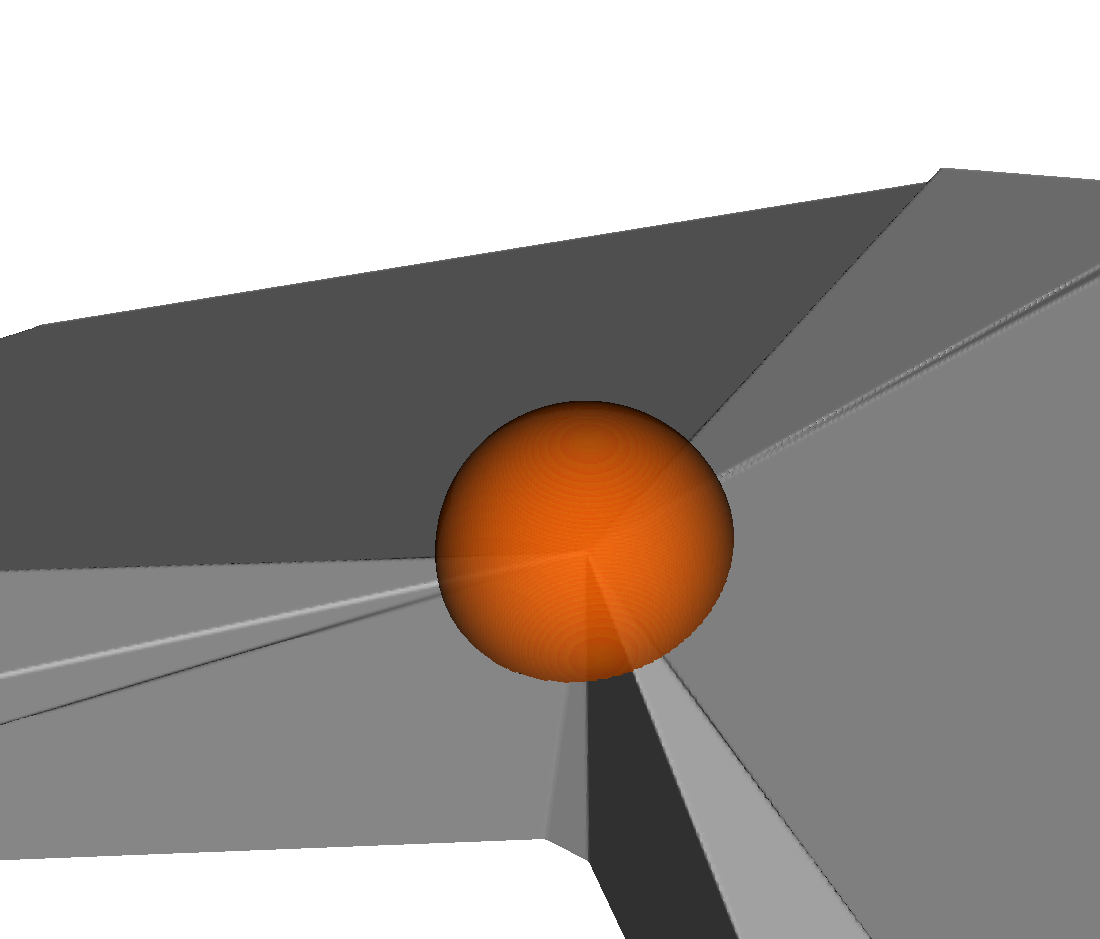}
\subcaption{SDF in cone}
\label{fig:pathological_3}
\end{subfigure}%
\begin{subfigure}[t]{0.25\textwidth}
\includegraphics[width=\textwidth]{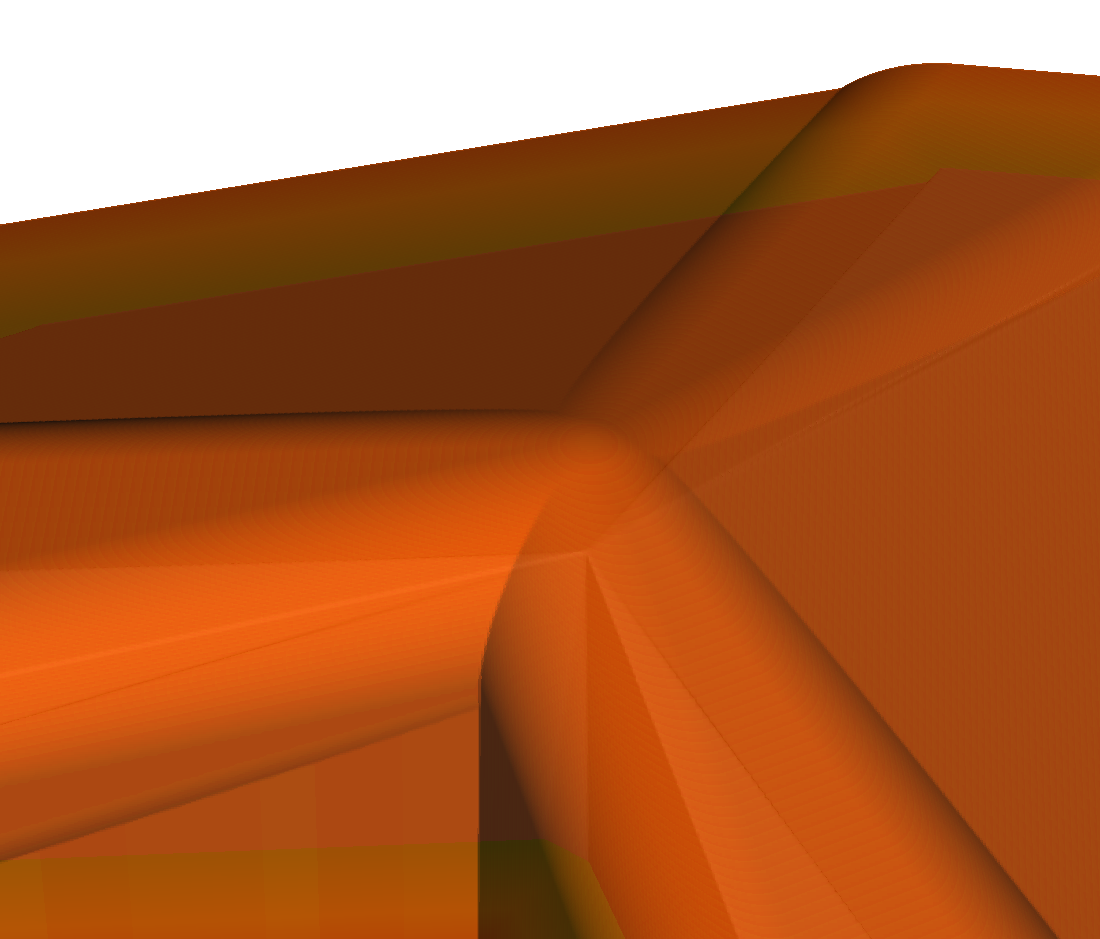}
\subcaption{Correct SDF}
\label{fig:pathological_4}
\end{subfigure}%
\caption{The pathological geometry case features a normal at a vertex which points in almost the opposite direction to the pseudonormal while the overall geometry is convex. Generating a cone of positive pointing normals fills the gap left between other extrusions.}
\label{fig:pathological}
\end{figure*}

While hemisphere generation at ruff geometries will produce the correct SDF, it emerged that it is sufficient to consider a cone extrusion restricted to positive pointing normals. Though the correctness of this approach is not certain, in all of the test cases, a cone encompassing just the positive pointing normals produced no gaps. The volume of such an extrusion is less than hemisphere and the workload is therefore smaller. Figure \ref{fig:ruff_fix} shows the SDF for the ruff geometry of figure \ref{fig:ruff}. The hole left at the convex vertex is filled by generating a cone enclosing the positive pointing normals. Following several attempts, no surface could be found which would lead to an incorrect SDF, although it is possible that such a configuration can occur in common geometries. Figure \ref{fig:pathological} shows a pathological test case which features a normal pointing almost in the negative pseudonormal direction with the vertex being categorised as convex (\ref{fig:pathological_1}). We show the hole left from other features (\ref{fig:pathological_2}), the SDF in the cone around positive pointing normals (\ref{fig:pathological_3}) and the correct distance field when applying the extrusion (\ref{fig:pathological_4}).


\begin{table}[h]
\begin{center}
\begin{tabularx}{0.55\textwidth}{|c *3{|>{\centering\arraybackslash}X}|}
\hline
Geometry & Faces & Time (s)\\
\hline 
Orion & 51,770 & 0.095\\
\hline 
Stanford Rabbit & 69,664 & 0.114 \\
\hline 
Stanford Dragon & 100,000 & 0.154\\
\hline 
XYZ RGB Dragon & 721,788 & 0.951\\
\hline 
Stanford Lucy & 2,529,647 & 3.105\\
\hline 
DrivAer & 2,854,762 & 3.601\\
\hline
\end{tabularx}
\caption{Internal geometry generation times for different STL files on K$20$ card.}
\label{tab:generation_k20}
\end{center}
\end{table}

\begin{table}[h]
\begin{center}
\begin{tabularx}{\textwidth}{|c *6{|>{\centering\arraybackslash}X}|}
\hline
Geometry & Total vertices & Unique & Saddle & Proportion\\
\hline 
Orion & 155,310 & 25,876 & 9,795 & 37.8\%\\
\hline 
Stanford Rabbit & 208,992 & 34,834 & 17,624 & 50.5\%\\
\hline 
Stanford Dragon & 300,000 & 50,000 & 26,431 & 52.8\%\\
\hline 
XYZ RGB Dragon & 2,165,364 & 360,894 & 192,882 & 53.4\%\\
\hline 
Stanford Lucy & 7,589,232 & 1,264,847 & 620,974 & 49.1\%\\
\hline 
DrivAer & 8,564,286 & 1,427,345 & 595,337 & 41.7\%\\
\hline
\end{tabularx}
\caption{The STL file format lists each vertex multiple times, and the resulting software combines them into unique points from which the appropriate extrusions are generated. For complex geometries, a large fraction of vertices are saddle points and need extrusions on both sides of the surface. The number of actual holes and errors in the SDF is different depending on the target domain resolution and the configuration of the surrounding surface.}
\label{tab:vertices_k20}
\end{center}
\end{table}

\subsection{Performance}

Table \ref{tab:generation_k20} shows the feature counts and generation times of internal geometry data for various bodies. We list the minimum recorded durations of several runs per shape. This includes reading in a binary STL file, generating entries on the CPU, copying them to the GPU where vertices and edges are sorted and combined into unique features. This timing also 
includes the construction of the extrusion polyhedra on the GPU. We note a stable scaling which depends heavily on the feature count. The timing also depends on the uniformity of triangle sizes and the extent of the STL geometry which determine the uniqueness of Morton codes and how much serialisation occurs in feature construction.

Table \ref{tab:vertices_k20} shows the number of vertices listed in the input STL file, how many unique points they are combined into and what the proportion of saddle points is. Not all unaddressed saddle vertices lead to visible errors in the SDF as surrounding extrusions may combine into watertight surfaces and depending on the resolution of the target mesh, the errors may not even be noticeable. However, the resulting SDF will not be accurate for every resolution and the likelihood of disruptive errors increases with the number of saddle points. For the complex test surfaces, the number of saddle points was between $37.8$\% and $53.4$\%, making it necessary to have a robust strategy to deal with high curvature vertices.


\begin{table}
\begin{center}
\setlength\doublerulesep{0.5pt}
\begin{subtable}{0.3\textwidth}
\centering
\begin{tabular}{| c| c| c|}
\hline 
\multirow{2}{*}{Distance} & \multicolumn{2}{c|}{Cell Size $\Delta x$} \\
\cline{2-3} 
& 0.08 & 0.04 \\
\hhline{|===|}
0.4  & 0.174 & 0.327 \\
\hline
0.8  & 0.180 & 0.486 \\
\hline
\end{tabular}
\caption{Orion}
\label{tab:sdf_time_orion_k20}
\end{subtable}
\begin{subtable}{0.3\textwidth}
\centering
\begin{tabular}{| c| c| c|}
\hline 
\multirow{2}{*}{Distance} & \multicolumn{2}{c|}{Cell Size $\Delta x$} \\
\cline{2-3} 
& 0.25 & 0.125 \\
\hhline{|===|}
2  & 0.223 & 0.242 \\
\hline
5  & 0.239 & 0.803 \\
\hline
\end{tabular}
\caption{Stanford Rabbit}
\label{tab:sdf_time_rabbit_k20}
\end{subtable}
\begin{subtable}{0.3\textwidth}
\centering
\begin{tabular}{| c| c| c|}
\hline 
\multirow{2}{*}{Distance} & \multicolumn{2}{c|}{Cell Size $\Delta x$} \\
\cline{2-3} 
& 0.16 & 0.08 \\
\hhline{|===|}
2  & 0.335 & 0.748 \\
\hline
5  & 1.090 & 7.192 \\
\hline
\end{tabular}
\caption{Stanford Dragon}
\label{tab:sdf_time_dragon_k20}
\end{subtable}

\begin{subtable}{0.3\textwidth}
\centering
\begin{tabular}{| c| c| c|}
\hline 
\multirow{2}{*}{Distance} & \multicolumn{2}{c|}{Cell Size $\Delta x$} \\
\cline{2-3} 
& 0.53 & 0.26 \\
\hhline{|===|}
5  & 2.214 & 2.266 \\
\hline
10 & 2.258 & 5.277 \\
\hline
\end{tabular}
\caption{XYZ RGB Dragon}
\label{tab:sdf_time_xyz_k20}
\end{subtable}\begin{subtable}{0.3\textwidth}
\centering
\begin{tabular}{| c| c| c|}
\hline 
\multirow{2}{*}{Distance} & \multicolumn{2}{c|}{Cell Size $\Delta x$} \\
\cline{2-3} 
& 4 & 2 \\
\hhline{|===|}
20 & 7.688 & 7.700 \\
\hline
40 & 7.721 & 7.741 \\
\hline
\end{tabular}
\caption{Stanford Lucy}
\label{tab:sdf_time_lucy_k20}
\end{subtable}
\begin{subtable}{0.3\textwidth}
\centering
\begin{tabular}{| c| c| c|}
\hline 
\multirow{2}{*}{Distance} & \multicolumn{2}{c|}{Cell Size $\Delta x$} \\
\cline{2-3} 
& 11.4e-3 & 5.7e-3 \\
\hhline{|===|}
0.06 & 8.535 & 8.513 \\
\hline
0.12 & 8.490 & 8.840 \\
\hline
\end{tabular}
\caption{DrivAer}
\label{tab:sdf_time_drivaer_k20}
\end{subtable}
\end{center}
\caption{SDF generation times in seconds for test geometries on K20 with dynamic parallelism.}
\label{tab:sdf_time_k20_dp}
\end{table}

Table \ref{tab:sdf_time_k20_dp} shows the time spent on generating the SDF for an Nvidia K$20$ card using dynamic parallelism. They list the minimum recorded durations of multiple runs. This includes kernel launches and tests if cells are within extrusions and writing appropriate values to global memory. The results show short generation time for the simpler test cases but also poor scaling for higher feature counts. Table \ref{tab:sdf_time_k20_ndp} shows the times when looping through bounding volume cells and not using dynamic parallelism. While the runtimes for simpler test cases are longer than for the parallel approach, as the feature count increases, the serial approach outperforms the alternative. This is due to both the higher launch cost of kernels on the device and a limited queue of active kernels and threads. The tipping point in performance is around $10^5$ faces, past which the serial approach is consistently better. An optimal implementation would then find a balance between maintaining the maximum amount of active parallel calculation and making sure the hardware queue is not oversubscribed by doing serial traversal of bonding volumes that may otherwise wait too long for device side launch.

Both table \ref{tab:sdf_time_k20_dp} and \ref{tab:sdf_time_k20_ndp} show how the runtime depends on the cell size of the domain and the maximum distance of the SDF. These variables are the main measures of workload for single bounding volumes. As the number of cells in the volumes increases, more points need to be tested for inclusion in the extrusions which means increased kernel launches or longer cell looping and potentially more conflicts in the atomic write to global memory. For the purposes of embedded mesh calculations a distance of only a couple of cells is needed to produce an accurate surface description which we can demonstrate short runtimes for.


\begin{table}
\begin{center}
\setlength\doublerulesep{0.5pt}
\begin{subtable}{0.3\textwidth}
\centering
\begin{tabular}{| c| c| c|}
\hline 
\multirow{2}{*}{Distance} & \multicolumn{2}{c|}{Cell Size $\Delta x$} \\
\cline{2-3} 
& 0.08 & 0.04 \\
\hhline{|===|}
0.4  & 0.234 & 1.745 \\
\hline
0.8  & 0.318 & 2.415 \\
\hline
\end{tabular}
\caption{Orion}
\label{tab:sdf_time_orion_k20_ndp}
\end{subtable}
\begin{subtable}{0.3\textwidth}
\centering
\begin{tabular}{| c| c| c|}
\hline 
\multirow{2}{*}{Distance} & \multicolumn{2}{c|}{Cell Size $\Delta x$} \\
\cline{2-3} 
& 0.25 & 0.125 \\
\hhline{|===|}
2  & 0.072 & 0.518 \\
\hline
5  & 0.257 & 1.941 \\
\hline
\end{tabular}
\caption{Stanford Rabbit}
\label{tab:sdf_time_rabbit_k20_ndp}
\end{subtable}
\begin{subtable}{0.3\textwidth}
\centering
\begin{tabular}{| c| c| c|}
\hline 
\multirow{2}{*}{Distance} & \multicolumn{2}{c|}{Cell Size $\Delta x$} \\
\cline{2-3} 
& 0.16 & 0.08 \\
\hhline{|===|}
2  & 0.123 & 0.866 \\
\hline
5  & 1.165 & 9.160 \\
\hline
\end{tabular}
\caption{Stanford Dragon}
\label{tab:sdf_time_dragon_k20_ndp}
\end{subtable}

\begin{subtable}{0.3\textwidth}
\centering
\begin{tabular}{| c| c| c|}
\hline 
\multirow{2}{*}{Distance} & \multicolumn{2}{c|}{Cell Size $\Delta x$} \\
\cline{2-3} 
& 0.53 & 0.26 \\
\hhline{|===|}
5  & 0.143 & 0.702 \\
\hline
10 & 0.591 & 3.972 \\
\hline
\end{tabular}
\caption{XYZ RGB Dragon}
\label{tab:sdf_time_xyz_k20_ndp}
\end{subtable}
\begin{subtable}{0.3\textwidth}
\centering
\begin{tabular}{| c| c| c|}
\hline 
\multirow{2}{*}{Distance} & \multicolumn{2}{c|}{Cell Size $\Delta x$} \\
\cline{2-3} 
& 4 & 2 \\
\hhline{|===|}
20 & 0.071 & 0.316 \\
\hline
40 & 1.395 & 1.548 \\
\hline
\end{tabular}
\caption{Stanford Lucy}
\label{tab:sdf_time_lucy_k20_ndp}
\end{subtable}
\begin{subtable}{0.3\textwidth}
\centering
\begin{tabular}{| c| c| c|}
\hline 
\multirow{2}{*}{Distance} & \multicolumn{2}{c|}{Cell Size $\Delta x$} \\
\cline{2-3} 
& 11.4e-3 & 5.7e-3 \\
\hhline{|===|}
0.06 & 0.078 & 0.339 \\
\hline
0.12 & 0.277 & 1.616 \\
\hline
\end{tabular}
\caption{DrivAer}
\label{tab:sdf_time_drivaer_k20_ndp}
\end{subtable}
\end{center}
\caption{SDF generation times in seconds for test geometries on K20 without dynamic parallelism.}
\label{tab:sdf_time_k20_ndp}
\end{table}

\subsection{Limitations}

While the current implementation introduces some improvements, there still remain limitations to the underlying algorithm. The CSC algorithm assumes a correct orientable surface, which means that there can be no flipped faces or gaps between faces. It is still possible to produce a correct SDF of a non-closed surface when clipping it to a smaller computational mesh where everything in the domain is either on one or the other side of the surface. The produced approach only creates a narrow band around the surface, leading to a secondary zero crossing between the negative limit of the SDF and the interior of the surface beyond the maximum distance. This can be easily fixed by sweeping along each of the coordinate axes and filling in unset values in the interior of the geometry. The geometry generation may be slow for large geometries with widely varying triangle sizes. In such domains, many smaller triangles can be assigned the same Morton code, leading to greater serialisation of the feature construction and longer generation times. This can be addressed by subdividing the input or distributing it across multiple cards.

\section{Conclusion}
\label{sec12}

Our work focused on describing embedded geometries in CFD simulations. These often feature relatively high resolutions and domains that extend far beyond the object surface. There is a need for quickly generating the SDF of complex geometries in limited regions of space, which still comprise a high number of small cells. The produced implementation allows for quick organisation and construction of internal geometry information, work scheduling and generating a signed distance field near object boundaries.

The original CSC algorithm has been adjusted to include angle weighted vertex normals and fixes for saddle points as seen in literature. We have also presented a discussion on problems of the original algorithm at high curvature vertices and a fix for these cases. A discussion on the nature of the extrusions has shown that there are no areas left uncovered by the union of extrusions and that sign conflicts do not lead to ambiguity. Though a hemisphere extrusion is the most certain way to ensure a correct SDF at high curvature vertices, in practice, a cone of positive pointing normals is sufficient. 

By using a set of common $3$D geometry test cases, we have shown the robustness of the algorithm and demonstrated the performance of both the geometry preparation as well as the SDF generation for a range of feature counts and domain resolutions. Like the original implementation of the algorithm, the performance scales with the feature count of the triangulated surface and the number of cells within the bounding volumes. 

We have presented a high performance generation of the necessary geometric data and the scheduling of work on GPUs. The resulting implementation offers a robust and fast way of generating $3$D signed distance fields in high resolution Cartesian grids.


\end{document}